\documentclass[conference]{IEEEtran}
\IEEEoverridecommandlockouts

\usepackage{graphicx} %
\usepackage{amsmath,amssymb}
\usepackage{xcolor}
\usepackage{url}
\usepackage{subcaption}
\usepackage{enumitem}
\usepackage[normalem]{ulem}
\usepackage{circledsteps}
\usepackage{booktabs}
\usepackage{balance}
\usepackage[hidelinks]{hyperref}

\AtBeginDocument{%
  }

\begin{document}
\bstctlcite{IEEEexample:BSTcontrol}

\title{PrivDiffuser: Privacy-Guided Diffusion Model for Data Obfuscation in Sensor Networks}

\author{\IEEEauthorblockN{Xin Yang}
\IEEEauthorblockA{\textit{University of Alberta} \\
Edmonton, Canada \\
xin.yang@ualberta.ca}
\and
\IEEEauthorblockN{Omid Ardakanian}
\IEEEauthorblockA{\textit{University of Alberta} \\
Edmonton, Canada \\
ardakanian@ualberta.ca}
}

\maketitle

\begin{abstract}
Sensor data collected by Internet of Things (IoT) devices can reveal sensitive personal information about individuals,
raising significant privacy concerns when shared with semi-trusted service providers,
as they may extract this information using machine learning models.
Data obfuscation empowered by generative models is a promising approach to 
generate synthetic data such that useful information contained in the original data is preserved
while sensitive information is obscured.
This newly generated data will then be shared with service providers instead of the original sensor data.
In this work, we propose PrivDiffuser, a novel data obfuscation technique 
based on a denoising diffusion model that achieves a superior trade-off between data utility and privacy
by incorporating effective guidance techniques.
Specifically, we extract latent representations that contain information about public and private attributes 
from sensor data to guide the diffusion model, and impose mutual information-based regularization 
when learning the latent representations to alleviate the entanglement of public and private attributes,
thereby increasing the effectiveness of guidance.
Evaluation on three real-world datasets containing different sensing modalities reveals
that PrivDiffuser yields a better privacy-utility trade-off than the state-of-the-art in data obfuscation, 
decreasing the utility loss by up to $1.81\%$ and the privacy loss by up to $3.42\%$.
Moreover, compared with existing obfuscation approaches, PrivDiffuser offers the unique benefit of allowing users with diverse privacy needs to protect their privacy without having to retrain the generative model.
\end{abstract}

\begin{IEEEkeywords}
Privacy-utility trade-off, Obfuscation, Deep generative models
\end{IEEEkeywords}

\section{Introduction}

The growing adoption of IoT devices brings data collection closer to our intimate spaces. 
Numerous IoT devices, such as smart home systems and wearables,
are equipped with cameras, microphones, and inertial measurement units~(IMUs),
enabling them to collect a wide range of data inconspicuously.
For device limitations and model security, 
these data are often sent to the cloud to make desired inferences using powerful deep learning models, 
e.g., to extract and analyze users' \emph{public attributes}.
While the service provider is assumed to faithfully perform the desired inferences, 
they may also perform unwanted inferences on these data, e.g., to extract users' \emph{private attributes}.
This honest-but-curious~(HBC) adversary might sell this sensitive information or use it for targeted advertisement.
For example, an individual's motion data collected for fitness tracking
can be used to infer their age, gender, or body size~\cite{hajihassnai2021obscurenet, malekzadeh2019mobile}.
Similarly, audio and video recorded by a baby camera for sleep monitoring can be used to infer the room layout and dimensions~\cite{Purushwalkam2021}.
These intrusive inferences pose a significant threat to our privacy, 
calling for the development of privacy-preserving techniques that are well-suited for IoT devices and applications.

\begin{figure}[t]
\centering
\includegraphics[width=\linewidth]{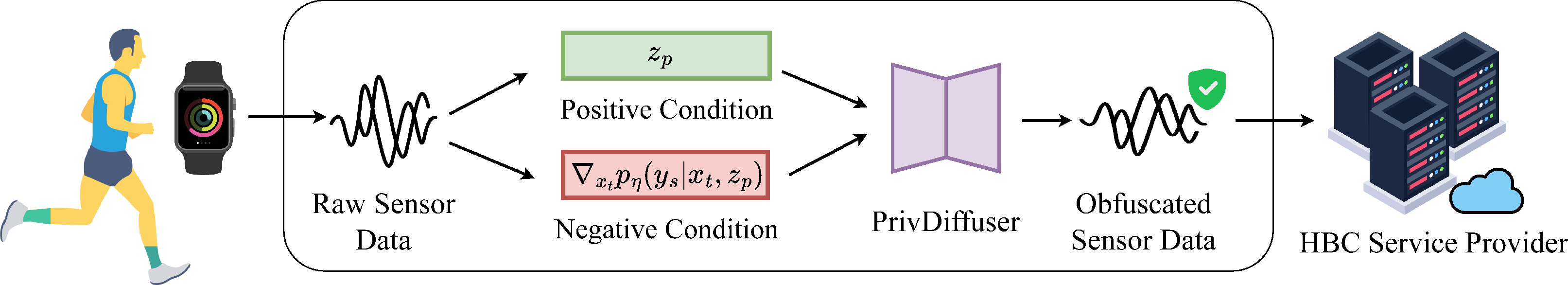}
\caption{The workflow of PrivDiffuser}
\label{fig:application_scenario}
\end{figure}

Although there are effective privacy-preserving techniques for structured data, 
sensor data presents unique challenges for privacy protection.
First, the private attributes
are usually embedded in time-series emitted by sensors but not directly present in the dataset,
making it difficult to mask or perturb them.
For example, a smartwatch is not equipped with a sensor for measuring weight, an attribute deemed private by some users,
but it integrates a variety of sensors for tracking different health metrics 
from which an individual's weight might be estimated.
In this case, weight measurements are not explicitly present in the dataset.
Second, the private attribute can be any attribute of the user 
and does not have to be their identity, which is normally protected in differential privacy~\cite{abadi2016deep}.
Existing privacy-preserving techniques based on Local Differential Privacy~(LDP)~\cite{erlingsson2014rappor,zheng2020privacy}
are not suitable for sensor data because adding calibrated noise directly 
to the sensor data could substantially reduce its utility. 
While it is possible to extract the private attribute from sensor data, 
add noise to that specific attribute, and then reconstruct the data using a decoder,
even a weak entanglement between public and private attributes in the latent space would 
void the privacy guarantees provided by differential privacy~\cite{weggenmann2022dp}.
The complete disentanglement of these attributes proves to be extremely difficult in some cases. %
The overall level of privacy is also determined by a fixed privacy budget in differential privacy,
which cannot be translated to a continual setting in a straightforward manner.
Existing cryptographic techniques for privacy protection, 
such as secure multi-party computation and homomorphic encryption~\cite{HCNN2021},
are computationally expensive for deep learning, and are unsuitable for resource-constrained IoT devices.

Privacy-preserving techniques based on generative models cast 
data obfuscation as a conditional generation problem,
i.e. the synthesized data must contain enough information for desired inferences (concerning public attributes), 
yet little sensitive information to prevent unwanted inferences (concerning private attributes)~\cite{raval2019olympus,malekzadeh2019mobile,hajihassnai2021obscurenet}.
Since sampling from a trained generative model is relatively inexpensive, obfuscation can be performed on IoT devices~\cite{yang2023blinder}.
Moreover, generative models produce new data in the same space as the input data, 
allowing downstream applications to seamlessly consume the obfuscated data.
For these reasons, obfuscation techniques that rely on generative models 
have received significant attention for real-world IoT applications.

In this work, we propose PrivDiffuser, a novel obfuscation technique
for sensor data based on a denoising diffusion model.
PrivDiffuser achieves a superior trade-off between utility and privacy compared to the previous work that 
integrates a conditional generative adversarial network~(GAN), 
and allows modifying the definition and importance of private attributes\footnote{Users sending data to a service provider may have different private attributes or prioritize privacy to varying degrees, and both of which can change over time.
However, their public attribute remains the same, as it is tied to the requested service.} at will, 
eliminating the need to retrain the diffusion model. 
Figure~\ref{fig:application_scenario} shows the workflow of PrivDiffuser.
In the forward process of the diffusion model, 
a small amount of Gaussian noise is added to the sensor data at each step, 
transforming it to near isotropic Gaussian noise after a number of steps. 
The backward process inverts this 
by training a machine learning model to predict the noise added at each step. 
This allows the diffusion model to generate realistic sensor data 
by progressively denoising randomly sampled Gaussian noise.

By \textit{guiding} the diffusion process using the information extracted from the original sensor data, 
we ensure that the newly generated data contains rich information about the public attributes (preserving utility)
and minimum information about the private attributes (mitigating privacy loss).
To achieve this, guidance is provided based on positive conditions for public attributes
and negative conditions for private attributes. 
Although classifier-free guidance~\cite{ho2022classifier} is more efficient than classifier guidance~\cite{dhariwal2021diffusion}
(since it uses classification labels rather than gradients derived from an auxiliary classifier),
the labels do not provide substantive information to reconstruct the obfuscated data.
Thus, we adopt classifier-free guidance to condition the diffusion model on the public attributes
using the latent features learned by a surrogate utility model.
For the private attributes, 
we design a negative conditioning technique based on classifier guidance and 
enhance it using forward universal guidance~\cite{bansal2023universal}.
This conditioning technique is suitable for data obfuscation as it allows 
guiding a trained diffusion model during the sampling stage, 
enabling users with diverse privacy needs to protect their private attributes with minimum retraining effort.
Lastly, the entanglement between the public and private attributes creates an intricate trade-off 
between utility and privacy. 
To maximally disentangle these attributes,
we impose mutual information-based regularization to extract a latent representation for each private attribute
that is %
weakly correlated with the 
public attributes, 
such that applying negative conditions to mitigate privacy loss does not compromise utility.
Our contribution is threefold:
\begin{itemize}[nosep]
    \item We propose a novel data obfuscation model based on a denoising diffusion model 
    and design effective guidance techniques for positive and negative conditions 
    governing data utility and privacy loss, respectively. 
    These guidance techniques offer the flexibility 
    to navigate the privacy-utility trade-off without retraining the diffusion model.

    \item We incorporate a mutual information-based regularization 
    to extract latent representations of private attributes 
    that are weakly correlated with public attributes
    to achieve a better privacy-utility trade-off.
    The private attribute representations are utilized to guide a trained diffusion model 
    without counteracting the effect of the positive conditions.

    \item We evaluate the proposed sensor data obfuscation model, PrivDiffuser, on three datasets to corroborate 
    its versatility and efficacy for different sensing modalities and private attribute definitions.
    Our results show that it outperforms the state-of-the-art GAN-based obfuscation model 
    with respect to privacy and utility, and can be easily extended to condition multiple public or private attributes.

\end{itemize}
We note that PrivDiffuser is intended to be deployed on the user's IoT device, ideally as a low-level module such that it has access to the raw sensor readings. This is necessary to ensure that third-party applications can only access the data obfuscated by PrivDiffuser.

\section{Related Work}
\subsection{Privacy-Aware Feature Extraction}

Privacy-aware feature extraction involves projecting sensor data into a subspace, producing a low-dimensional representation that contains the maximum amount of information
about the public attribute(s) and almost no information about the private attribute(s).
Previous work in this area combines deep learning and adversarial training.
For example, Privacy Adversarial Network (PAN)~\cite{liu2019privacy} and DeepObfuscator~\cite{li2021deepobfuscator} train an encoder alongside a classification model for the target task
and two adversarial networks aiming to reconstruct the original data
and predict the private attribute(s) respectively. 
Similarly, Li \textit{et~al.}~\cite{li2020tiprdc} used mutual information between random variables to train a feature extractor that minimizes privacy loss while maintaining data utility
through adversarial training.
The common drawback of these techniques is that
the learned representation is not in the same space as the original sensor data.
Thus, it cannot be readily consumed by existing applications that take sensor data as input, 
requiring substantial changes to the apps, 
which is laborious and often impractical.

\subsection{Data Obfuscation via Generative Models} 
Data obfuscation can be performed using a sequence of transformations that result in an output
sharing the same dimensions as the input data, 
allowing seamless integration with existing applications.
Given the input sensor data, 
a generative model can produce new data by obscuring the sensitive information contained in the original data
while keeping the useful information for the target inference task. 
If the generative process is conditioned properly, 
this approach leads to a reasonable trade-off between data utility and privacy loss.
DoppelGANger~\cite{lin2020using} shows the possibility of synthesizing high-fidelity time-series data using GANs 
and recurrent neural networks.
But it does not obscure sensitive information in the generated data.
An autoencoder-based data obfuscation model is proposed in~\cite{malekzadeh2019mobile},
where the autoencoder is regularized using multiple loss terms 
to reduce sensitive information in its latent space.
Olympus~\cite{raval2019olympus} uses adversarial training to 
iteratively train an autoencoder-based obfuscation model and an attacker model. 
This allows minimizing privacy loss while maintaining the utility of data.
ObscureNet~\cite{hajihassnai2021obscurenet} uses a conditional variational autoencoder (CVAE) for data obfuscation. 
A discriminator network classifies the private attribute 
given the latent representation extracted by the encoder/generator.
The discriminator is jointly trained with the CVAE in an adversarial fashion to encourage the CVAE to obscure sensitive information in the latent space.
Recently, Chen \textit{et~al.} proposed MaSS~\cite{chen2024mass}, a data transformation model based on information theoretic measures that suppresses multiple sensitive private attributes. 
The authors assume the availability of two types of public attributes, annotated and unannotated, which is different from our setting.
Moreover, the evaluation metric defined in that paper considers any intrusive inference accuracy below the random guessing level as equally effective as random guessing.
In summary, existing data obfuscation models are mostly built on the autoencoder architecture, 
where the encoder acts as a generator and is jointly trained with a discriminator in the same spirit as GAN. 
For this reason, we classify them as GAN-based obfuscation models.

\begin{figure}[!t]
\centering
\includegraphics[width=\linewidth]{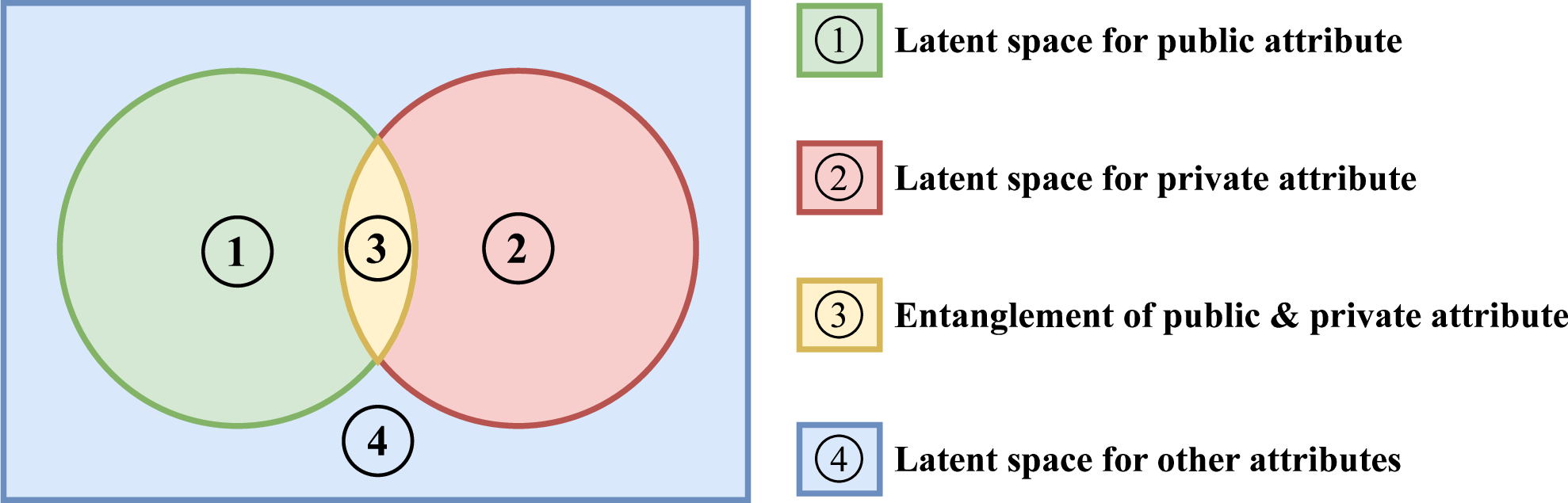}
\caption{GAN-based obfuscation uses a discriminator to hide information about the private attribute in \textcircled{\raisebox{-0.9pt}{2}}+\textcircled{\raisebox{-0.9pt}{3}}.
PrivDiffuser is conditioned on the public attribute to include information in \textcircled{\raisebox{-0.9pt}{1}}+\textcircled{\raisebox{-0.9pt}{3}} and the negated private attribute to exclude private information in \textcircled{\raisebox{-0.9pt}{3}}. 
The white-listing characteristic of diffusion-based obfuscation
offers protection for the private attribute and all attributes that were not listed by the user, 
extending privacy protection to \textcircled{\raisebox{-0.9pt}{2}}+\textcircled{\raisebox{-0.9pt}{3}}+\textcircled{\raisebox{-0.9pt}{4}} (see Appendix~\ref{subsec:whitelisting_characteristic}).
For ease of presentation, the entanglement between the public/private attribute and other attributes is not shown.}
\label{fig:latent}
\end{figure}

Diffusion models~\cite{sohl2015deep} are
another type of generative models that have gained popularity due to their outstanding performance 
in generating high-quality, realistic images or videos. %
Ho \textit{et~al.} proposed Denoising Diffusion Probabilistic Model~(DDPM)~\cite{ho2020denoising} 
by modeling the diffusion process as a Markov chain. 
In DDPM, the sampling (denoising) process can be slow
because generating a sample requires synthesizing data in all previous steps.
To accelerate the sampling process, 
Song \textit{et~al.}~\cite{song2020denoising} proposed Denoising Diffusion Implicit Model~(DDIM). DDIM offers a non-Markovian perspective on the diffusion process, enabling faster sampling.
A recent study~\cite{yang2023privacy} suggests that the denoising diffusion model is suitable for data obfuscation
if the diffusion model is conditioned on the public attribute only,
as other attributes will be sampled randomly from a diverse distribution learned from the training data. 
This \textit{white-listing} approach ensures that only information about the public attribute
is embedded in the obfuscated data, as illustrated in Figure~\ref{fig:latent}.
However, since attributes are entangled in the latent space, 
some information about the other attributes will be included in the obfuscated data inevitably.
This problem was not addressed in~\cite{yang2023privacy}.
That said, the denoising diffusion model is indeed a promising approach for data obfuscation, 
offering several advantages over GAN-based obfuscation.
Specifically, training a diffusion model does not entail a min-max game between a generator and a discriminator,
hence it is more stable and efficient.
Further, in GAN-based obfuscation, 
different privacy-utility trade-offs can be achieved by tuning the weight of the discriminator loss,
but this requires retraining the obfuscation model.
In contrast,
we build PrivDiffuser on top of a diffusion model, making it possible to achieve different trade-offs
by tuning parameters at the sampling stage,
without retraining the obfuscation model.
More details are provided in Section~\ref{sec:composition} and~\ref{subsec:tuning_tradeoff}.

\subsection{Conditional Diffusion Models}
The output of an unconditional diffusion model is highly stochastic due to the randomness 
in the backward diffusion process.
Extensive research has been done in recent years 
on conditioning the diffusion model and guiding the sampling process
so as to control the content of the generated data~\cite{dhariwal2021diffusion,ho2022classifier,nichol2021glide}.
In particular, Dhariwal \textit{et~al.}~\cite{dhariwal2021diffusion} 
proposed \emph{classifier guidance}, a guidance technique that uses the gradient of an auxiliary classifier 
to condition the diffusion model on the target class information.
Since the auxiliary classifier takes as input the noisy data generated by the diffusion model, 
training it under various noise levels is extremely difficult.
Bansal \textit{et~al.} proposed \emph{universal guidance}~\cite{bansal2023universal},
a guidance technique that substitutes the noisy input of the auxiliary classifier during sampling with predicted clean data, making it possible to train the auxiliary classifier on clean data for improved performance.
Ho \textit{et~al.}~\cite{ho2022classifier} proposed \emph{classifier-free guidance}, 
a guidance technique that directly conditions the diffusion model using the (one-hot encoded) label of the target class.
Considering that the label itself has no semantic structure and 
labels in multiple conditions might be conflicting,
the latent representation learned by a semantic encoder is used to condition a DDIM in~\cite{preechakul2022diffusion}.
Other studies utilized pre-trained image-language models, 
such as contrastive language-image pre-training~\cite{radford2021learning}, 
to guide the image generation process using text prompts~\cite{nichol2021glide,kim2022diffusionclip}.
Further, diffusion models have achieved state-of-the-art performance in image inpainting tasks 
through conditioning on the masked images~\cite{rombach2022high}. 
Yet image inpainting~\cite{suvorov2022resolution} has fundamental differences with data obfuscation 
because the unwanted content in data obfuscation is the information that is not explicitly included in the dataset 
but can be inferred from the sensor data.
The above guidance techniques only consider a positive condition that 
the diffusion model should include in the generated data. 

In many applications, it is crucial to also guide the diffusion model with negative conditions, 
such that specific content is removed from the synthesized data.
Du \textit{et~al.}~\cite{du2020compositional} and Liu \textit{et~al.}~\cite{liu2022compositional} 
interpreted diffusion models as energy-based models
and proposed the composition of multiple text conditions 
through conjunction and negation operations for visual generation.
The conjunction operation enables the composition of multiple positive conditions, 
whereas the negation operation allows the composition of negative conditions.
These negation techniques require pre-training a series of diffusion models,
each conditioned on one of the respective conditions. 
This drastically increases the training cost.  
Similar approaches have also been used in text-to-image models 
based on latent diffusion models~\cite{gandikota2023erasing, dong2024towards}.
Armandpour \textit{et~al.}~\cite{armandpour2023re} found that the composition of complex text input conditions
requires conditional independence between positive and negative conditions,
which is not often the case in the real world.
To address this, the authors proposed the Perp-Neg technique
to compute the overlapped semantics between the denoising score components of the positive and negative conditions 
and then subtract it from the denoising score component of the negative condition. 
However, we found empirically that Perp-Neg is not very effective for sensor data obfuscation.
An alternative approach to disentangle the positive and negative conditions
is to project the representations of these conditions onto the orthogonal space of each other. 
For instance, Deng \textit{et~al.}~\cite{deng2023fairness} proposed extracting latent representations of the public attribute
that are orthogonal to the subspace of the private attribute, 
where the latent representations for the public and private attributes are extracted using dedicated encoders. 
This approach requires knowledge of the private attribute prior to the training of the public attribute encoder, 
which is not ideal when the private attribute can change over time,
because the diffusion-based data obfuscation model must be retrained in that case. 
In our work, we propose a mutual information-based regularization approach 
for extracting latent representations of the private attribute.
This enables post-training guidance based on the negative condition 
using the latent representations of the private attribute,
such that 
PrivDiffuser
can be reused 
to protect different private attributes without retraining. 
We expand on this in Section~\ref{subsec:negative_guidance}.

\noindent \textbf{Novelty of this work:} 
We put forward an innovative approach to data obfuscation based on the diffusion process
that is well-suited for IoT systems and achieves a superior privacy-utility trade-off 
by alleviating the entanglement of public and private attributes in a latent space.
Our work differs from the previous work on conditional denoising diffusion models in three ways.
First, existing methods are mostly designed for text-to-image generation tasks, 
whereas our work takes multi-channel sensor data segments (multivariate time-series) as input and utilizes a diffusion model to obscure the sensitive information contained in sensor data.
Second, to guide text-to-image generation, off-the-shelf models, such as Word2Vec~\cite{church2017word2vec},
are used to extract word embeddings from the available text prompts.
But, in data obfuscation, public and private attributes must be inferred and transformed, 
e.g., projected into a latent space, before they are used to condition the diffusion model.
Third, 
we incorporate a mutual information-based regularization 
to apply the negative condition without counteracting the effect of the positive condition,
leading to a better privacy-utility trade-off.

\section{Background on Conditional Denoising Diffusion Model}

\subsection{Training}
\label{subsec:background_training}
A diffusion model consists of a forward diffusion process and a backward diffusion process.
The forward process is divided into $T$ timesteps,
where Gaussian noise is added to the data in each timestep. 
We denote the original data as $x_0$, 
its latent probability distribution as $q$, %
and the perturbed data at timestep $t$ as $x_t$.
The amount of noise added at each timestep $t$ is modeled using a variance scheduler $\beta_t$, where $0{<}\beta_1 {<}\cdots{<}\beta_T{<}1$. 
In DDPM, the diffusion process is modeled as a Markov process, such that
the noisy data at timestep $t$ of the forward process 
depends on the data at the previous timestep $t-1$ as follows:
\begin{equation}
\label{eq:ddpm_xt}
    p(x_t|x_{t-1}) = \mathcal{N}(\sqrt{1-\beta_t}~x_{t-1}, \beta_t \mathrm{I}).
\end{equation}
The downside of using the above relation is that 
the denoised data at every timestep must be computed and stored in the sampling process. 
This can be addressed by rewriting~(\ref{eq:ddpm_xt}) such that $x_t$ depends on $x_0$ only.
Specifically, by introducing $\alpha_t = 1-\beta_t$ and $\bar{\alpha} = \prod_{t=1}^T \alpha_t$, we can write:
\begin{equation}
    \label{eq:ddpm_x0}
    p(x_t|x_0) = \mathcal{N}\big(\sqrt{\bar{\alpha}} x_0, (1-\bar{\alpha}) \mathrm{I}\big),
\end{equation}
and its reparameterization yields 
$
    x_t = \sqrt{\bar{\alpha}}x_0 + \sqrt{1-\bar{\alpha}}\epsilon,
$
where $\epsilon$ follows a standard Gaussian distribution, i.e. $\epsilon{\sim}\mathcal{N}(\mathbf{0}, \mathbf{I})$. 
Notice that for sufficiently large $T$, $x_T$ will have nearly an isotropic Gaussian distribution. 
This offers an intuitive interpretation of the diffusion model: 
\emph{novel data ($x_0$) can be generated
by sampling $x_T$ from the standard Gaussian distribution
and progressively denoising it using the reverse process.}
To this end, the intractable reverse process $q(x_{t-1}|x_t)$ 
must be approximated using a neural network model $p_\theta$.
When the noise schedule $\beta$ is sufficiently small, 
$p_\theta(x_{t-1}|x_t)$ can also be considered to have a Gaussian distribution:
\begin{equation}
    p_\theta(x_{t-1}|x_t) = \mathcal{N}\big(\mu_\theta(x_t,t), \Sigma_\theta(x_t,t)\big),
\end{equation}
where $\mu_\theta$ and $\Sigma_\theta$ are its mean and variance, respectively.
Following Ho \textit{et~al.}~\cite{ho2020denoising}, 
we use a time-dependent constant ($\beta_t^2$) for the variance to simplify training, 
and train a neural network to predict the noise ($\epsilon$) that is
added to data rather than 
a neural network predicting the mean of the Gaussian distribution $p_\theta$.
Let $\epsilon_\theta(x_t)$ be the model that predicts $\epsilon$.
This noise-prediction network is trained using the mean squared error (MSE) 
between the actual noise and the predicted noise:
\begin{equation}
\label{eq:mse}
    \mathcal{L}_\theta = \|\epsilon - \epsilon_\theta(x_t, t)\|_2^2 = \| \epsilon - \epsilon_\theta(\sqrt{\bar{\alpha}}x_0 + \sqrt{1-\bar{\alpha}} \epsilon, t) \|_2^2,
\end{equation}
where %
$\epsilon$ is the noise randomly drawn from $\mathcal{N}(\textbf{0},\mathbf{I})$,
and $t$ is the timestep uniformly sampled between $1$ and $T$. 
We use the UNet architecture~\cite{ronneberger2015u} for $\epsilon_\theta(x_t)$. 
We note that the noise-prediction network can be used to derive a \textit{score function} 
as outlined in~\cite{dhariwal2021diffusion}:
\begin{equation}
\label{eq:score}
    \nabla_{x_t} \log p_\theta(x_t) = -\frac{1}{\sqrt{1-\bar{a}_t}} \epsilon_\theta(x_t).
\end{equation}

To condition the diffusion model, 
the condition $c$ 
is introduced in the backward diffusion process. 
Concretely, the noise prediction model receives $c$ as input, 
so it is expressed as $\epsilon_\theta(x_t, t, c)$. 
Similar to an unconditional diffusion model,
the predicted noise is compared with the noise introduced in the forward diffusion process and the following MSE loss is optimized to train this model~\cite{ho2020denoising}:
\begin{equation}
\label{eq:mse_condition}
    \mathcal{L}_\theta = 
    \| \epsilon - \epsilon_\theta(\sqrt{\bar{\alpha}}x_0 + \sqrt{1-\bar{\alpha}} \epsilon, t, c) \|_2^2.
\end{equation}
We discuss how to encode an attribute to get $c$ in Section~\ref{subsec:positive_guidance}.

\subsection{Sampling}
We briefly explain how to guide the diffusion model via classifier guidance and classifier-free guidance 
during sampling in DDIM~\cite{song2020denoising}.

\subsubsection{Classifier Guidance~\cite{dhariwal2021diffusion}}

In this approach,
the condition is the output label ($y$) of an auxiliary classifier parameterized by $\phi$.
This classifier is trained to predict $y$ given the input data $x_t$.
The score function for $p(x_t)p(y|x_t)$ is given by:
\begin{equation}
\label{eq:classifier_guidance_ddim}
    \begin{aligned}
        \nabla_{x_t}\!\log\big(p_\theta(x_t)p_\phi(y|x_t)\big)\!=\!\nabla_{x_t}\log p_\theta(x_t) + \nabla_{x_t}\!\log p_\phi(y|x_t) \\
        =\!-\frac{\epsilon_\theta(x_t){-}\sqrt{1-\bar{a}_t}\nabla_{x_t}\!\log p_\phi(y|x_t)}{\sqrt{1-\bar{a}_t}},
    \end{aligned}
\end{equation}
where the last derivation is obtained by using~(\ref{eq:score}).
We now define a new epsilon prediction $\bar{\epsilon}(x_t)$ 
corresponding to the score of the joint distribution:
\begin{equation}
\label{eq:classifier_guidance_ddim_epsilon}
    \bar{\epsilon}(x_t) :=\epsilon_\theta(x_t) - \sqrt{1-\bar{a}_t} \nabla_{x_t} \log p_\phi(y|x_t).
\end{equation}
This modified noise prediction $\bar{\epsilon}(x_t)$ is used instead of $\epsilon_\theta(x_t)$ during sampling 
to repeatedly predict $x_{t-1}$ given $x_t$, until we get $x_0$~\cite{dhariwal2021diffusion}:
\begin{equation}
\label{eq:predict_xt}
    x_{t-1} = \sqrt{\bar{\alpha}_{t-1}}\Big(\frac{x_t-\sqrt{1-\bar{\alpha}_{t}}\bar{\epsilon}(x_t)}{\sqrt{\bar{\alpha}_{t}}}\Big) + \sqrt{1-\bar{\alpha}_{t-1}}\bar{\epsilon}(x_t).
\end{equation}

\textit{Remark.} Since $\bar{\epsilon}(x_t)$ depends on the unconditional diffusion model
rather than the conditional diffusion model, 
it is not necessary to train $\epsilon_\theta(x_t,t,y)$ for this approach
and it suffices to introduce the condition at sampling time.
This is beneficial because changing the condition does not require retraining the diffusion model.
We expand on this in Section~\ref{sec:method}.

\subsubsection{Classifier-free Guidance~\cite{ho2022classifier}}
Classifier guidance requires training an additional classifier to obtain its gradient 
$\nabla_{x_t}\log p(y|x_t)$
for sampling.
In classifier-free guidance, an \textit{implicit classifier} is obtained 
by jointly training a conditional and an unconditional diffusion model.
Let the unconditional denoising diffusion model $p_\theta(x)$
be parameterized through a score estimator $\epsilon_\theta(x_t,t)$
and the conditional model $p_\theta(x|y)$ through $\epsilon_\theta(x_t,t,y)$,
with $y$ being the one-hot encoding %
of a given class label. 
These two models can be obtained by training a single neural network because $\epsilon_\theta(x_t,t){=}\epsilon_\theta(x_t,t,y{=}\varnothing
)$, where $y{=}\varnothing$ indicates that no bit is set in $y$.

Considering the following relation:
\begin{equation}
\label{eq:joint-model-training}
     p(y|x_t)\propto \frac{p(x_t|y)}{p(x_t)},
\end{equation}
the score function of $p_\theta(y|x)$ is given by:
\begin{equation}
\label{eq:classifier_free_score}
    \begin{aligned}
        \nabla_{x_t}\log p_\theta(y|x) &= \nabla_{x_t}\log p_\theta(x_t|y) - \nabla_{x_t}\log p_\theta(x_t) \\
        &= -\frac{1}{\sqrt{1-\bar{a}_t}} 
        \big(
        \epsilon_\theta(x_t,t,y) - \epsilon_\theta(x_t,t)
        \big).
    \end{aligned}
\end{equation}
By plugging (\ref{eq:classifier_free_score}) into (\ref{eq:classifier_guidance_ddim_epsilon}) 
and introducing the scale term $w$,
a hyperparameter that governs the importance of guidance, %
we have:
 \begin{equation}
     \begin{aligned}
          &\bar{\epsilon}_\theta(x_t,t,y) = \epsilon_\theta(x_t,t,y) {-} \sqrt{1-\bar{a}_t} w \nabla_{x_t}\log p_\theta(y|x_t) \\
          &= \epsilon_\theta(x_t,t,y) {-} \sqrt{1-\bar{a}_t} w \big(  \nabla_{x_t}\log p_\theta(x_t|y) {-} \nabla_{x_t}\log p_\theta(x_t)  \big) \\
        &= (1+w) \epsilon_\theta(x_t,t,y) - w \epsilon_\theta(x_t,t).
     \end{aligned}
\end{equation}
This $\bar{\epsilon}_\theta(x_t,t,y)$ is used in the sampling process as discussed in~\cite{ho2022classifier}.

\textit{Remark.} Since $\bar{\epsilon}_\theta(x_t,t,y)$ depends on the conditional diffusion model,
the condition must be introduced at the training stage 
so that it can affect the sampling process.

\section{Guiding Diffusion Sampling Towards a Desirable Privacy-Utility Trade-off}
\label{sec:method}
\begin{figure}[t]
\centering
\includegraphics[width=\linewidth]{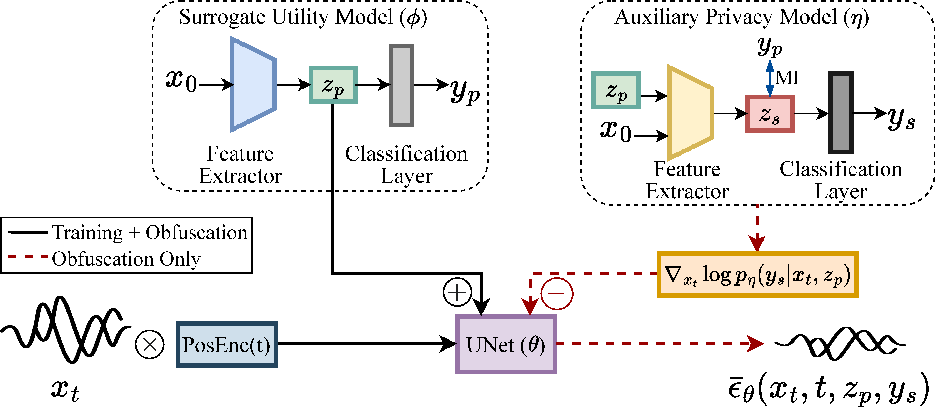}
\caption{Architecture of PrivDiffuser}
\label{fig:architecture}
\end{figure}

We propose obfuscating sensor data via the iterative denoising procedure of a diffusion model
that is guided to add information about the true public attribute label
while ensuring that no information about the true private attribute label will be added in this process.
Below we describe our threat model and outline the specific techniques that are employed in the training and sampling stages
of PrivDiffuser to achieve a good trade-off between utility and privacy.
We first discuss how to encode the public and private attributes and use them to condition the diffusion model.
We then introduce an effective regularization technique 
to alleviate the entanglement between these attributes.
Finally, we explain how to adapt the training process when the user specifies multiple public or private attributes.
These techniques are shown in Figure~\ref{fig:architecture}.
We denote the labels for the public attribute and the private attribute
as $y_p$ and $y_s$, and the latent representation of $y_p$ and $y_s$ extracted 
by the feature extractors as $z_p$ and $z_s$, respectively. 
$\eta$ is the parameter of the auxiliary privacy model
predicting $y_s$, %
and $\phi$ is the parameter of the surrogate utility model predicting $y_p$. %
These models are described in Section~\ref{subsec:positive_guidance} and~\ref{subsec:negative_guidance}.

\subsection{Threat Model}
We consider an HBC adversary that has access to sensor data $x$ shared by users to faithfully perform desired inferences about their public attributes $y_p$. 
Being curious, the adversary wishes to infer certain private attributes $y_s$ that are not explicitly disclosed from the shared sensor data, via an Attribute Inference Attack (AIA). 
We assume that the adversary has 
no metadata about the users who shared their sensor data,
does not know if the shared data was obfuscated,
and performs the desired and intrusive inferences on a remote server using models that are not exposed to users.
These inference models can be trained on a public dataset that contains sensor data and metadata (i.e. corresponding attributes).
PrivDiffuser's objective is to mitigate such AIAs by transforming raw sensor data $x$ into $x'$, in a way that ensures the desired inferences can be performed accurately, i.e. $P(y_p|x'){\approx} P(y_p|x)$, while reducing the adversary's ability to infer private attributes to the level of random guessing i.e., $P(y_s|x'){\approx} P_\text{random guess}$.
PrivDiffuser is assumed to be capable of replacing raw data with obfuscated data on the user device, before sharing the data with the HBC adversary.
It does not require knowledge of the intrusive and desired inference models used by the adversary and uses surrogate models instead.

\subsection{Classifier-free Guidance for Enhanced Data Utility}
\label{subsec:positive_guidance}
Classifier-free guidance offers better sampling quality than classifier guidance by training a diffusion model with both conditioned and unconditioned data~\cite{ho2022classifier}.
To exploit the simplicity and efficiency of classifier-free guidance 
while tackling the unique challenges in sensor data obfuscation,
we propose training a classifier to predict the public attribute 
and conditioning the diffusion model with the latent representation extracted by 
the first few layers of this classifier (i.e., the feature extractor part) rather than its output. 
Intuitively, the learned latent representation carries 
richer semantics about the public attribute
compared to the one-hot encoded public attribute label.
Thus, using this representation for classifier-free guidance makes the obfuscated data more similar to real data.

The public attribute classifier
serves as our \textit{surrogate utility model}.
It takes a raw sensor data segment $x$ as input and 
predicts the public attribute label $y_p$. %
Note that the actual model used in the target task 
to predict the public attribute
may differ from the surrogate utility model, hence the name.
For instance, the obfuscated data could eventually be processed on the server side 
by a more sophisticated or deeper neural network 
than the surrogate utility model.

We adopt a convolutional neural network as our surrogate utility model. 
This network consists of 2 convolutional layers followed by 4 fully connected (FC) layers.
The convolutional layers and the first 3 FC layers constitute the feature extractor.
The first FC layer takes as input the latent representations learned by the convolutional layers and 
compresses the latent representation to a size of 256. 
The second and third FC layers then scale down the latent representations using 128, and 60 neurons, respectively. 
The gradual decrease in the size of the latent space forces the surrogate utility model 
to condense important features that are highly correlated to the public attribute and remove other information.
We use ReLU as the activation function and normalize the output of the third FC layer. 
The output of the feature extractor, denoted as $z_p$, is incorporated
as the condition $c$ in the diffusion model as described in Section~\ref{subsec:background_training}.
We then stack the fourth FC layer after the feature extractor 
and use the Softmax activation function to map $z_p$ into a probability distribution 
over public attribute classes. 
The cross entropy loss is used for training the surrogate utility model.

Given a trained surrogate utility model, 
we incorporate the learned representation of the public attribute 
into the training process of the diffusion model.
Specifically, adaptive group normalization~(AdaGN)~\cite{dhariwal2021diffusion} is used
to condition the diffusion model on both the timestep $t$ and the latent representation $z_p$:
\begin{equation}
\label{eq:adagn}
    \text{AdaGN}(h,y)=y_t \text{GroupNorm}(h)+y_{z_p},
\end{equation}
where $h$ is the output of the first convolutional layer in the UNet's residual block, $y_t$ is the linear transformation of the positional encoding~\cite{vaswani2017attention} of the timestep $t$, $y_{z_p}$ is the linear transformation of the latent representation $z_p$.
Finally, $y:=[y_t, y_{z_p}]$ is incorporated as a condition into the UNet.

Once trained, the conditioned diffusion model is used in the sampling process
to ensure that the obfuscated data will contain information about the same public attribute 
as the original sensor data. 
We emphasize that the surrogate utility model is trained before training the diffusion model. 
This way, during sampling, the surrogate utility model is simply used to obtain the representation 
of the public attribute as we do not need to calculate its gradient for classifier-free guidance.
As a result, sampling is less expensive in this case 
compared to the classifier guidance approach where the gradient of the classifier must be computed for sampling.

\subsection{Privacy Protection via Classifier Guidance} %
\label{subsec:negative_guidance}
To minimize privacy loss, sampling must be done
such that sensitive information about the private attribute is not included in the obfuscated data.
To this end, we incorporate a negative condition for each private attribute 
using the classifier guidance approach, and call this classifier the \textit{auxiliary privacy model}.
A negative condition enforces the \textit{absence} of information 
pertaining to a particular label in the generated data, 
whereas a positive condition, which is used for a public attribute, 
enforces the \textit{presence} of information pertaining to a particular label in the generated data.
The primary reason for conditioning the private attributes via classifier guidance is that
unlike classifier-free guidance that requires training a conditional diffusion model,
classifier guidance can be directly applied at sampling time. 
This characteristic allows users with diverse privacy needs to reuse the same pre-trained diffusion model 
conditioned on a specific public attribute and eliminates the complexity of training a conditional diffusion model 
that takes into account all potential combinations of the conditions.

\subsubsection{Combining Positive and Negative Conditions to Achieve a Desirable Privacy-Utility Trade-off}
\label{sec:composition}
We extend the derivations for sampling using classifier guidance 
to the case that there are both positive and negative conditions
for public and private attributes, respectively. 
The negative condition is denoted as $\bar{y}_s$, which represents the complement of $y_s$ or $\textbf{not}(y_s)$.
Our goal is to compute $p_{\theta,\phi,\eta}(x_t| y_p, \bar{y}_s)$.
Considering a realistic scenario in which $y_p$ and $y_s$ are entangled, i.e., they are not conditionally independent,
we can write~\cite{dong2024towards,liu2022compositional}:
\begin{equation}
    \begin{aligned}
    \label{eq:proportional}
    p_{\theta,\phi,\eta}(x_t| y_p, \bar{y}_s) &\propto 
        p_{\theta,\phi,\eta}(x_t, y_p, \bar{y}_s) \\
        &\propto p_\theta(x_t)\frac{p_\phi(y_p|x_t)}{p_\eta(y_s|y_p, x_t)}.
    \end{aligned}
\end{equation}
Taking the gradient of the logarithm of (\ref{eq:proportional}) w.r.t. $x$ yields:
    \begin{align}
    &\nabla_{x_t} \log\big(p_\theta(x_t)p_\phi(y_p|x_t) p_\eta(\bar{y}_s|, y_p, x_t)\big) =\\
    \nonumber &\nabla_{x_t} \log p_\theta(x_t) {+} \nabla_{x_t} \log p_\phi(y_p|x_t){-}\nabla_{x_t} \log p_\eta(y_s|y_p, x_t) = \\
    \nonumber &-\frac{1}{\sqrt{1-\bar{a}_t}}\epsilon_\theta(x_t)
        {+} \nabla_{x_t}\log p_\phi(y_p|x_t) {-} \nabla_{x_t}\log p_\eta(y_s|y_p, x_t),
\end{align}
where in the last step we substituted the noise prediction model for the score function using (\ref{eq:score}).
Finally, we define $\bar{\epsilon}_\theta(x_t,t,y_p,\bar{y}_s)$ that is used in the sampling process:
\begin{equation}
\label{eq:black_list_classifier_guidance_ddim_epsilon}
\begin{aligned}
    \bar{\epsilon}_\theta(x_t,t,y_p,\bar{y}_s)
    = \epsilon_\theta(x_t) - \sqrt{1-\bar{a}_t}\nabla_{x_t}\log p_\phi(y_p|x_t) \\ + \sqrt{1-\bar{a}_t}\nabla_{x_t}\log p_\eta(y_s|y_p, x_t).
\end{aligned}
\end{equation}

Recall that we use an implicit classifier for the public attribute 
because classifier-free guidance is used in that case.
Thus, we substitute $-\frac{1}{\sqrt{1-\bar{a}_t}} \big(\epsilon_\theta(x_t,t,y_p) - \epsilon_\theta(x_t,t)\big)$ 
for $\nabla_{x_t}\log p_\phi(y_p|x_t)$ according to~(\ref{eq:classifier_free_score}),
and use the latent representation $z_p$ instead of $y_p$ (as discussed in Section~\ref{subsec:positive_guidance})
to update the noise-prediction network:
\begin{equation}
\label{eq:negation_loss_ddim}
\begin{aligned}
        \bar{\epsilon}_\theta(x_t,t,z_p,y_s) =
        (1+w_1) \epsilon_\theta(x_t,t,z_p) - w_1 \epsilon_\theta(x_t,t) \\ + w_2 \sqrt{1-\bar{a}_t} \nabla_{x_t}\log p_\eta(y_s|z_p, x_t).
\end{aligned}
\end{equation}
The auxiliary privacy model $p_\eta(y_s|y_p, x_t)$ shares the same architecture as the surrogate utility model $p_\phi(y_p|x_t)$, except that for the auxiliary privacy model, we concatenate the public latent representation $z_p$ generated by $p_\phi(y_p|x_t)$ with the flattened output of the second convolutional layer 
in the feature extractor of 
the auxiliary privacy model
and feed the result to its first FC layer.
Here $w_1$ and $w_2$ are hyperparameters that are introduced 
to control the influence of positive and negative conditions in the sampling (obfuscation) process.
By tuning these hyperparameters, one can navigate the utility-privacy trade-off
as discussed in Section~\ref{subsec:tuning_tradeoff}.

\subsubsection{Regularizing Auxiliary Privacy Model via Mutual Information}
The proposed approach for combining positive and negative conditions
relaxes the strong assumption that the corresponding attributes are conditionally independent.
In this setting, guiding the diffusion model toward generating obfuscated data that does not contain information about the private attribute could remove information about the public attribute, 
worsening the privacy-utility trade-off.
For example, an individual's height might be correlated with their race, 
so one cannot simply combine these conditions %
to generate sensor data that belongs to a tall person while hiding their race.
Thus, to achieve a desirable privacy-utility trade-off, 
we encourage learning a latent representation of the private attribute, $z_s$, that is weakly correlated %
with the 
latent representation of the public attribute, $y_p$, 
such that introducing the negative condition will not counteract the effect of the positive condition.

We propose regularizing the auxiliary privacy model
by minimizing the mutual information (MI) between $z_s$ and the public attribute label $y_p$.
This is accomplished by augmenting the loss function of the auxiliary privacy model,
which will in turn affect the representation learned for the private attribute.
Since computing the exact MI between $z_s$ and $y_p$ is intractable, we estimate the MI between $z_s$ and $y_p$ by using the Mutual Information Neural Estimator (MINE)~\cite{belghazi2018mutual}, a neural network that is trained to predict the approximate MI given $z_s$ and $y_p$. 
We choose the one-hot encoded public attribute label $y_p$ rather than the latent representation $z_p$ because although both correlate with the public attribute label, 
$z_p$ has a larger dimension than $y_p$ and is learned from the original sensor data $x$, hence it can contain extraneous information apart from information about the public attribute.
Thus, using $y_p$ in MINE is more efficient than $z_p$ and yields better disentanglement.

We implement MINE using a neural network parameterized by $\mu$ with four fully connected layers where each layer is followed by a ReLU activation function.
The first layer takes as input the concatenation of the two random variables and use 400 neurons to compute the output. 
The second and third FC layers also use 400 neurons, and the last FC layer outputs a single value 
estimating the MI between the two input random variables.
We train MINE together with the auxiliary privacy model. 
In particular, for every mini-batch of training data that contains $b$ data samples, 
we first obtain the private latent representation $z_s$ by feeding the clean data input $x_0$ 
and the public attribute representation $z_p$ to the privacy auxiliary classifier $p_\eta(y_s|z_p, x_0)$, 
where $z_p$ is computed using the pre-trained feature extractor in the surrogate utility model.
Next, MINE is trained to estimate MI between $z_s$ and $y_p$ following the algorithm proposed in~\cite{belghazi2018mutual}.
We use Adam optimizer with a learning rate of 0.0002 to train MINE.
Then $z_p$ is fed into the classification layer of the auxiliary privacy model to predict the private attribute label.
We regularize the auxiliary privacy model by using MINE to estimate the MI between $z_s$ and $y_p$:
\begin{equation}
\label{eq:privacy_loss}
    \mathcal{L}_\eta = \mathcal{L}_{\text{CE}} + w_3 \cdot
    {\text{MI}(z_s, y_p)},
\end{equation}
where $\mathcal{L}_{\text{CE}}$ is the cross-entropy loss for predicting the private attribute label, and $w_3$ is a hyperparameter that controls the strength of the regularization term. Note that although $w_3$ can also serve as a tunable knob to navigate the privacy-utility trade-off, we argue that it is more intuitive and preferable to only tune $w_1$ and $w_2$ with a fixed $w_3$.
We empirically select a value of $w_3$ that encourages disentangling the public and private attributes to the maximum extent while preserving high accuracy of the auxiliary privacy model.

\subsubsection{Improving Effectiveness of Negative Conditions using Universal Guidance}
The negative conditioning is achieved via classifier guidance 
to take advantage of its ability to guide the diffusion model during the sampling stage,
where the input of the auxiliary privacy model is the data perturbed with noise that is sampled at a random timestep $t$. This requires $p_\eta$ to achieve high accuracy on noisy input $x_t$.
However, we found that training $p_\eta$ on the noisy data causes convergence issues 
and reduces the effectiveness of negative conditioning.
Therefore, we train $p_\eta(y_s|y_p, x_t)$ on clean sensor data and follow the idea of universal guidance~\cite{bansal2023universal} to substitute the noisy input $x_t$ with the predicted clean data $\hat{x}_0$ 
during sampling.
Specifically, the denoised data $x_{t-1}$ is sampled from $q(x_{t-1}|x_t, \hat{x}_0)$ in DDIM, where 
$\hat{x}_0=\frac{x_t-\sqrt{1-\bar{\alpha}_{t}}\bar{\epsilon}(x_t)}{\sqrt{\bar{\alpha}_{t}}}$
is the predicted denoised data that substitutes the unknown $x_0$,
as shown in~(\ref{eq:predict_xt}). 
Since the diffusion model is conditioned on the public attribute in our work,
we apply the forward universal guidance to a conditional diffusion model by substituting $x_t$ with:
\begin{equation}
    \begin{aligned}
    \label{eq:predicted_clean_data}
        \hat{x}_0=&\frac{x_t-\sqrt{1-\bar{\alpha}_{t}}\bar{\epsilon}_\theta(x_t, t, z_p)}{\sqrt{\bar{\alpha}_{t}}} \\
        =& \frac{x_t-\sqrt{1-\bar{\alpha}_{t}} \big( (1+w_1) \epsilon_\theta(x_t,t,z_p) - w_1 \epsilon_\theta(x_t,t) \big)}{\sqrt{\bar{\alpha}_{t}}}.
    \end{aligned}
\end{equation}
Finally, we update our noise-prediction network as follows:
\begin{equation}
\begin{aligned}
\label{eq:negation_loss_ddim_universal}
         \bar{\epsilon}_\theta(x_t,t,z_p,y_s) =
         (1+w_1) \epsilon_\theta(x_t,t,z_p) - w_1 \epsilon_\theta(x_t,t) \\ + w_2 \sqrt{1-\bar{a}_t} \nabla_{x_t}\log p_\eta(y_s|z_p, \hat{x}_0).
\end{aligned}
\end{equation}

\subsection{Extension to Multiple Private Attributes and Multiple Public Attributes}
We now explain how multiple private attributes can be protected simultaneously 
via a straightforward extension of the proposed negative conditioning technique.
Consider a total of $K$ private attributes denoted as $y_{s_1}, y_{s_2}, {\cdots}, y_{s_K}$. 
For each private attribute $k \in [1,2,{\cdots}, K]$, we adopt an auxiliary privacy model parameterized by $\eta_k$.
Our goal is to compute $p_{\theta, \phi, \eta_1, {\cdots}, \eta_K}(x_t|y_p, \bar{y}_{s_1},{\cdots}, \bar{y}_{s_K})$.
For the ease of exposition,
we derive the noise predictor by assuming that the $K$ private attributes are conditionally independent, 
given $x_t$ and the public attribute $y_p$.\footnote{
This assumption simplifies the optimization problem and the dependency between private attributes is expected to result in stronger privacy protection at the cost of slightly reducing utility. That being said, one can disentangle the private attributes first before performing negative conditioning to improve the privacy-utility trade-off.}
Similar to (\ref{eq:proportional}), we can write:
\begin{equation}
    \begin{aligned}
    \label{eq:multi_private}
    & p_{\theta,\phi,\eta_1, {\cdots}, \eta_K}(x_t| y_p, \bar{y}_{s_1}, {\cdots}, \bar{y}_{s_K}) {\propto}  \\
    & p_{\theta,\phi,\eta_1, {\cdots}, \eta_K}(x_t, y_p, \bar{y}_{s_1},{ \cdots}, \bar{y}_{s_K}) 
        {\propto} \frac{p_\theta(x_t) p_\phi(y_p|x_t)}{\prod_{k=1}^Kp_{\eta_k}(y_{s_k}|y_p, x_t)}.
    \end{aligned}
\end{equation}
Taking the gradient of the logarithm of (\ref{eq:multi_private}) w.r.t. $x$ yields:

\begin{align}
    \nonumber &\nabla_{x_t} \log p_\theta(x_t) {+} \nabla_{x_t}\!\log p_\phi(y_p|x_t){-}\!\sum_{k=1}^K \nabla_{x_t}\!\log p_{\eta_k}(y_{s_k}|y_p, x_t) \\
    &{=}{-}\frac{\epsilon_\theta(x_t)}{\sqrt{1{-}\bar{a}_t}}
    {+} \nabla_{x_t}\!\log p_\phi(y_p|x_t) {-}\!\sum_{k=1}^K \nabla_{x_t}\!\log p_{\eta_k}(y_{s_k}|y_p, x_t).
\end{align}
The rest of the derivation follows (\ref{eq:black_list_classifier_guidance_ddim_epsilon}) and (\ref{eq:negation_loss_ddim}). 
We can write the noise predictor $\bar{\epsilon}$ as:
\begin{equation}
\begin{aligned}
    &\bar{\epsilon}_\theta(x_t,t,z_p,y_{s_1},{\cdots},y_{s_K}) {=} (1+w_1) \epsilon_\theta(x_t,t,z_p) {-} w_1 \epsilon_\theta(x_t,t) \\ 
    & {+} \sum_{k=1}^K w_{k+1} \sqrt{1-\bar{a}_t} \nabla_{x_t}\log p_\eta(y_{s_k}|z_p, x_t),
\end{aligned}
\end{equation}
where $w_{k+1}$ is a hyperparameter that controls the strength of negative conditioning for private attribute $y_{s_k}$.
The mutual information-based regularization technique and the improved classifier guidance based on forward universal guidance 
can also be trivially applied to train each auxiliary privacy model.

It is also possible to support desired inferences about multiple public attributes. 
We use multi-task learning as an example to demonstrate the feasibility of conditioning $K$ public attributes, 
denoted $y_{p_1}$, $\cdots$, $y_{p_K}$. 
In particular, we first modify the surrogate utility model $p_\phi$ to predict multiple public attributes simultaneously. Specifically, the latent representation, $z_p$, extracted by the feature extractor of the surrogate utility model is used to separately predict $y_{p_1}$, $\cdots$, $y_{p_K}$ via $K$ classification heads. 
Each classification head uses the cross-entropy (CE) loss to measure the error of predicting its corresponding public attribute. The $K$ CE loss terms are then summed to obtain the total classification loss of the surrogate utility model, which is trained by using the Adam optimizer, following the same approach introduced in Section~\ref{subsec:positive_guidance}.
We also adapt training of the auxiliary privacy model to disentangle the private attribute from the public attributes. 
To this end, for simplicity, we consider the case of protecting one private attribute $y_s$ and employ $K$ MINE models, 
each estimating the mutual information between the latent representation of the private attribute $z_s$ and the label of a public attribute. This encourages the auxiliary privacy model to learn private attribute information that is weakly correlated with the public attributes. Hence, we can modify the loss function of the auxiliary privacy model in (\ref{eq:privacy_loss}) as:
\begin{equation}
\label{eq:privacy_loss_2pub}
    \mathcal{L}_\eta = \mathcal{L}_{\text{CE}} + \sum_{k=1}^K w_{p_k} \cdot
    {\text{MI}(z_s, y_{p_k})},
\end{equation}
where $w_{p_k}$ controls the strength of the MI-based regularization for disentangling private attribute with the public attribute $y_{p_k}$.
In Appendix~\ref{appendix:B}, we analyze the scalability of PrivDiffuser by conditioning on up to four attributes (3:1 and 1:3 for public:private attributes).

\section{Evaluation}
\label{sec:evaluation}

\subsection{Baselines}
\subsubsection*{ObscureNet~\cite{hajihassnai2021obscurenet}}
ObscureNet is a GAN-based data obfuscation model that utilizes a CVAE
jointly optimized with a discriminator in an adversarial fashion.
The discriminator encourages the encoder of CVAE to learn latent representations that do not contain sensitive information. 
By conditioning the decoder of CVAE with a %
dummy private attribute label, this obfuscation model generates sensor data that contain misleading private information to reduce the privacy loss while maintaining the utility.
We use ObscureNet as a representative GAN-based obfuscation model because its implementation is publicly available, unlike~\cite{raval2019olympus} and some other related work, and it is shown to outperform strong GAN-based baselines in terms of the privacy-utility trade-off~\cite{hajihassnai2021obscurenet}.
Note that, in ObscureNet, a dedicated obfuscation model is trained per public attribute class, whereas our approach trains a single diffusion model for all public and private attributes. 
Thus it has a higher training cost than PrivDiffuser.
ObscureNet is trained and evaluated using the code released by the authors on GitHub. %
We use the randomized anonymization approach, 
as it is robust to the re-identification attack.

\subsubsection*{Diffusion}
Our next baseline is an obfuscation model based on 
a denoising diffusion model with the same architecture as PrivDiffuser. 
We use the same conditioning approach for the public attribute, but we do not apply the negative condition
for the private attribute(s).
This baseline 
highlights the white-listing capability of diffusion-based obfuscation models that only use the positive condition for the public attribute.

\subsubsection*{Diffusion with Negation}
We implement this baseline on top of the \emph{Diffusion} baseline by adding the proposed negative conditioning technique to enforce the absence of information about the private attribute 
in the obfuscated data.
However, we do not apply the proposed mutual information-based regularization to alleviate the entanglement between public and private attributes.
This baseline is used to establish two things:
First, the effectiveness of the negative conditioning technique
through comparison with the \emph{Diffusion} baseline; 
Second, the role of MI-based regularization
in disentangling public and private attributes through
comparison with PrivDiffuser.

\subsection{Datasets}
Our evaluation is conducted on three publicly available datasets
including different modalities that are commonly used for human activity recognition~(HAR).
The first two datasets use IMU sensors, which are increasingly embedded in mobile and IoT devices, and the third one uses WiFi signals.

\subsubsection*{MobiAct~\cite{chatzaki2016human}}
It is an activity recognition dataset collected using the accelerometer, gyroscope, and orientation sensor in a Samsung smartphone~\cite{chatzaki2016human}.
It contains data from 66 participants 
performing 12 different daily activities.
For a fair comparison with our first baseline~\cite{hajihassnai2021obscurenet} 
that used a subset of this dataset, 
we use only the accelerometer and gyroscope data 
to detect these 4 activities:
walking, standing, jogging, and walking up the stairs.
This yields 6 channels from the 3-axis accelerometer 
and the 3-axis gyroscope.
We select the same 36 participants (20 male and 16 female participants) used in that baseline.
We preprocess the data by performing standardization and segmentation
using a sliding window of size 128 samples and a stride length of 10 samples.
We split the data with an 8:2 train-test ratio.
The public attribute that participants wish to infer from their sensor data is \textit{activity}. 
We consider two private attributes that were recorded for each participant: \textit{gender} and \textit{weight}.
Gender is a binary attribute in this dataset, 
and weight is considered as
a ternary attribute comprising Group 0 (${\leq}$ 70kg), Group 1 (70-90kg), and Group 2 (${\geq}$ 90kg).
This models individuals' weight as a categorical attribute containing three groups of nearly the same size.

\subsubsection*{MotionSense~\cite{malekzadeh2019mobile}}
It is collected using the accelerometer and gyroscope sensor of an iPhone~6s, carried by 24 participants 
(14 male and 10 female participants).
For a fair comparison with our first baseline~\cite{hajihassnai2021obscurenet},
we consider the following 4 activities 
from the 6 activities recorded in the dataset: 
walking up and down the stairs, walking, and jogging.
Each activity is repeated in 15 trials. 
We use the first 9 trials for training and the remaining 6 trials for testing, similar to the baseline. 
The public attribute is \textit{activity} and 
the private attribute is the user's \textit{gender},
a binary attribute in this dataset.
Magnitude readings from the accelerometer and gyroscope are aggregated over their respective 3 axes, 
resulting in 2 channels, one for each sensor. 
The aggregated sensor data is then segmented using the same technique used for the MobiAct dataset.

\subsubsection*{WiFi-HAR~\cite{baha2020dataset}}
It uses radio frequency~(RF), in particular WiFi channel state information~(CSI), for activity recognition. 
The WiFi-HAR dataset contains 90 CSI channels collected from 3 transceiver pairs (1 transmitter and 3 receivers), with each transceiver pair generating 30 channels.
We follow~\cite{yang2023blinder} and use the data collected in a line-of-sight environment, 
where 10 participants performed 12 activities. 
Each activity is repeated 20 times.
We select 4 activities, namely standing, sitting, lying down, and turning around. This constitutes our public attribute.
The participants' \textit{weight} is chosen as the private attribute 
and categorized into two groups: higher and lower than 80kg.
To preprocess the data, the magnitude of the CSI readings for each channel is computed and standardized. We then apply a sliding window of 80 samples with a stride length of 40 samples to segment the data. 
The segments are randomly shuffled and divided into training and test sets with a split ratio of 8:2.

\subsection{Evaluation Metrics}
\label{subsec:eval_metrics}
\subsubsection{Privacy Loss}
Our first evaluation metric is the success rate of AIA, 
which signifies the loss of private information 
due to sharing data with an HBC service provider.
To measure the success rate of AIA, 
we use the 
classification accuracy of a powerful intrusive inference model predicting the private attribute given the obfuscated data--a standard practice in the data obfuscation literature~\cite{chen2024mass,hajihassnai2021obscurenet,li2020tiprdc,li2021deepobfuscator,liu2019privacy,malekzadeh2019mobile,raval2019olympus,yang2023blinder}.
This model is a convolutional neural network~(CNN) that has $4$ convolutional layers, 
followed by $3$ fully connected layers.
We note that an adversary can use any arbitrary model. In this work, we adopt this CNN architecture as it simulates a powerful attacker by using a more advanced architecture compared to the auxiliary privacy model and surrogate utility model, and it achieves high inference accuracy on raw sensor data.
We train this model on raw sensor data,
and use its classification accuracy as a measure of privacy loss.
The training and test sets are resampled and different from the ones used to train PrivDiffuser.
Ideally, the classification accuracy on obfuscated data
should be close to the level of random guessing 
(e.g., $50\%$ for binary private attribute) instead of $0\%$. 
This is because the latter is prone to a re-identification attack, where the adversary learns 
the quasi-deterministic mapping and recovers the original private attribute label~\cite{hajihassnai2021obscurenet}.

\subsubsection{Data Utility}
We use the classification accuracy of a powerful desired inference model as a measure of data utility.
This inference model takes as input the obfuscated sensor data and predicts the public attribute label.
It is a more sophisticated model
than the surrogate utility model used in the obfuscation pipeline, showing that data obfuscated using a simpler model can maintain high utility.
We construct the desired inference model using the same architecture as the intrusive inference model and train it on the raw sensor data.
Overall, the obfuscation model should not decrease the utility of data, meaning that the desired inference accuracy should remain the same as the case that raw sensor data is used.

\begin{table*}[hbt!]
    \centering
    \resizebox{\linewidth}{!}{
    \begin{tabular}{lcccccccccc}
        \toprule
        \textbf{Activity} 
        & \multicolumn{2}{c}{Walking} & \multicolumn{2}{c}{Standing} & \multicolumn{2}{c}{Jogging} & \multicolumn{2}{c}{Upstairs} & \multicolumn{2}{c}{\textbf{Overall} (Accuracy/F1-score)} \\
        \cmidrule(r){1-1} \cmidrule(r){2-3} \cmidrule(r){4-5} \cmidrule(r){6-7} \cmidrule(r){8-9} \cmidrule(r){10-11}
        \textbf{Private Attribute}  & Gender & Weight & Gender & Weight & Gender & Weight & Gender & Weight & Gender & Weight \\
        \midrule
        Raw Data & \multicolumn{2}{c}{98.10\%} & \multicolumn{2}{c}{99.58\%} & \multicolumn{2}{c}{99.74\%} & \multicolumn{2}{c}{95.14\%} & \multicolumn{2}{c}{98.80\% / 91.85\% } \\
        ObscureNet & 94.82\% & 94.96\% & 99.60\% & 99.08\% & 98.14\% & 97.78\% & 93.00\% & 92.04\% & 97.21\% / 84.99\% & 97.02\% / 84.29\% \\
        Diffusion & \multicolumn{2}{c}{96.84\%} & \multicolumn{2}{c}{99.19\%} & \multicolumn{2}{c}{99.40\%} & \multicolumn{2}{c}{87.86\%} & \multicolumn{2}{c}{97.96\% / 87.44\% }  \\
        Diffusion with Negation & 95.74\% & 95.08\% & 98.30\% & 94.62\% & 99.64\% & 99.42\% & 89.72\% & 83.46\% & 97.07\% / 84.54\% & 94.93\% / 79.53\% \\
        PrivDiffuser & 95.96\% & 96.12\% & 98.78\% & 97.92\% & 99.62\% & 99.46\% & 90.80\% & 86.76\% & 97.40\% / 85.49\% & 97.03\% / 84.17\% \\
        \bottomrule
    \end{tabular}
    }
    \caption{Average activity recognition accuracy of obfuscation models on the MobiAct dataset. Note that the Diffusion baseline does not assume the knowledge of the private attribute, hence the HAR accuracy for gender/weight obfuscation is the same.
    }
    \label{tab:utility_mobi_both}
\end{table*}
\subsection{Implementation Details}\label{sec:impl}
PrivDiffuser is trained on a GPU server and deployed on edge/IoT devices to perform on-device data obfuscation.
We implement the surrogate utility model and auxiliary privacy model using a simple CNN that contains 2 convolutional layers followed by 4 FC layers as introduced in Section~\ref{subsec:positive_guidance}. 
The third FC layer outputs the latent feature representations of size $60$, and the fourth FC layer is the classification layer. 
We use ReLU as the activation function.
For MobiAct, MotionSense, and WiFi-HAR datasets, we set the number of output channels of the (first, second) convolutional layers to (8, 16), (64, 128), and (16, 32), respectively. 
Both convolutional layers use a kernel size of 2. 
We implement the noise prediction network, $\epsilon_\theta$, 
based on an open-source UNet architecture released by OpenAI~\cite{unet-git}. 
We use a linear scheduler~\cite{ho2020denoising} 
that starts from 0.0001 and ends at 0.02 with $T{=}1000$.
A total of 50 steps are used in the sampling stage to generate obfuscated data. 
We use a batch size of 128 for MobiAct and MotionSense, and a smaller batch size of 8 for WiFi-HAR for GPU memory efficiency.
In this setting, obfuscating one batch containing 128 sensor data segments from the MobiAct dataset takes approximately $11$ seconds on one NVIDIA RTX 2080 Ti GPU, that is approximately $86$ ms for obfuscating one sensor data segment (see Appendix~\ref{appendix:B} for details). 
We find that there is a linear relationship between the number of steps used in the sampling stage and the obfuscation time.
For MobiAct, we set the number of model channels used in the UNet to 64, 
and hyperparameters as follows: $w_1{=}2.5$, $w_2{=}0.5$, $w_3{=}8$ for gender obfuscation, and 
$w_1{=}3.8$, $w_2{=}1.5$, and $w_3{=}8$ for weight obfuscation.
For MotionSense, we set the number of model channels to 256, $w_1{=}7.8$, $w_2{=}0.05$, and $w_3{=}8$. 
For WiFi-HAR, we set the number of model channels to 256, $w_1{=}5.8$, $w_2{=}0.4$, and $w_3{=}4$.
The values of $w_1$ and $w_2$ are chosen so as to get a near-random guessing level of accuracy from the
intrusive inference model. 
This makes the diffusion-based models robust to re-identification attacks~\cite{hajihassnai2021obscurenet}.
We use the same hyperparameters for the two diffusion-based baselines. 
Note that for the Diffusion baseline, we set $w_1{=}3.8$ for both gender and weight obfuscation in MobiAct, 
since it does not use the negative conditioning technique.
Our implementation is released on GitHub: {\url{https://github.com/sustainable-computing/PrivDiffuser}}.

\section{Case Studies}
\subsection{Case Study 1: HAR on MobiAct}
\label{subsec:mobiact}

\subsubsection{Protecting Gender}
\begin{figure}[!t]
    \centering
    \begin{subfigure}[b]{0.49\linewidth}
         \centering
         \includegraphics[width=\linewidth]{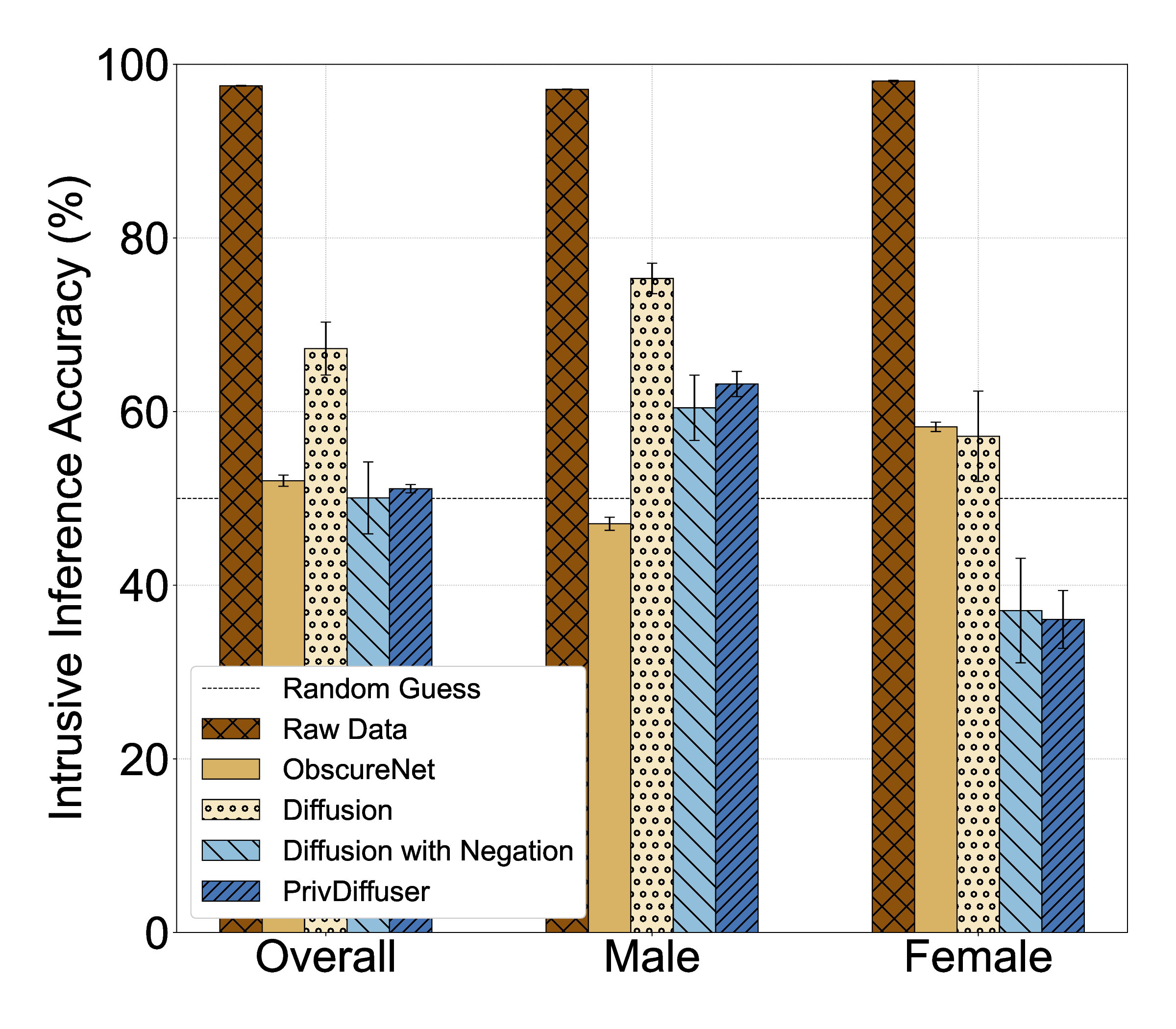}
         \caption{Gender obfuscation}
         \label{fig:privacy_mobi_gen}
     \end{subfigure}
     \hfill
    \begin{subfigure}[b]{0.49\linewidth}
         \centering
         \includegraphics[width=\linewidth]{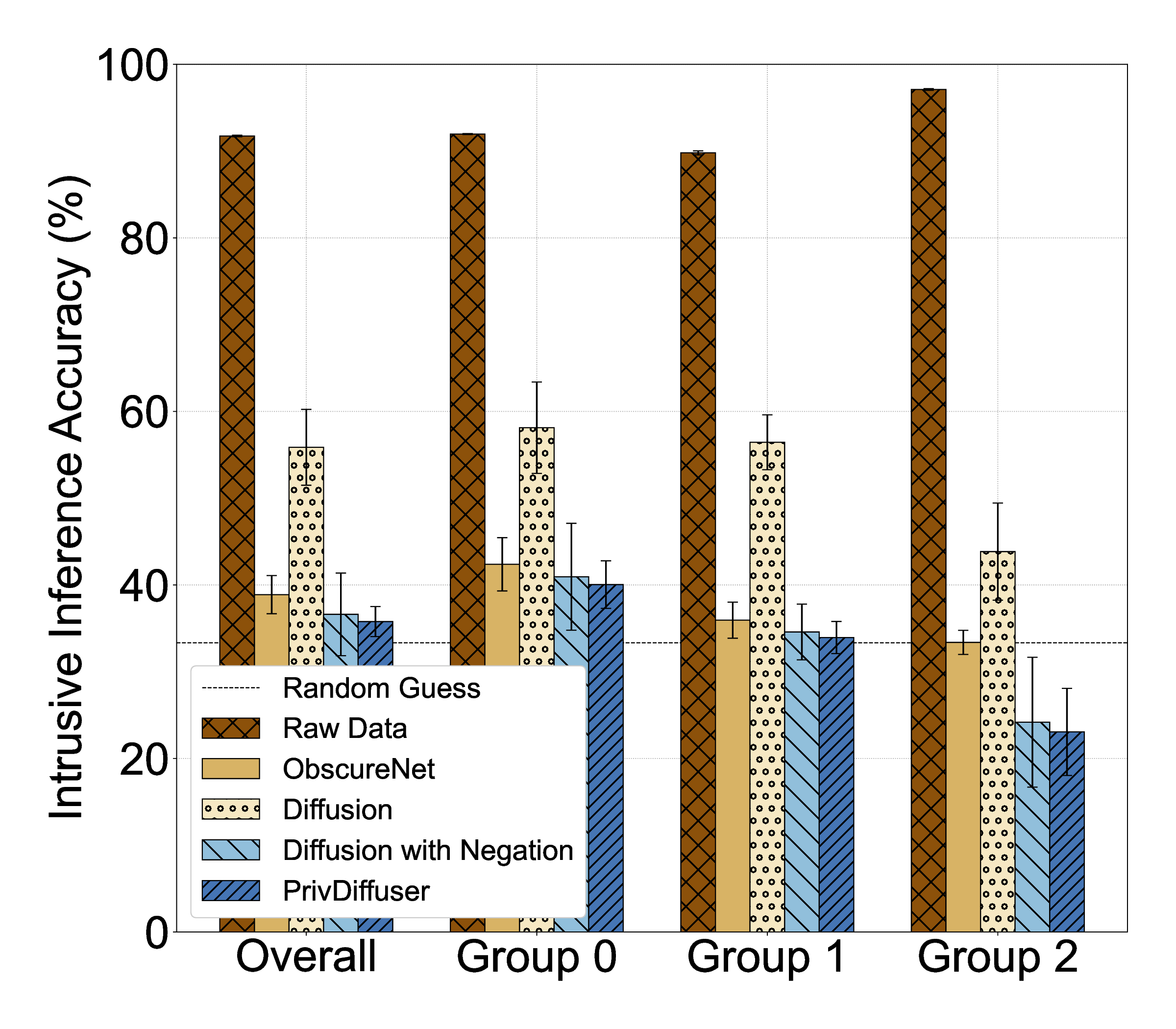}
         \caption{Weight obfuscation}
         \label{fig:privacy_mobi_weight}
    \end{subfigure}
    \caption{Intrusive inference accuracy on MobiAct dataset for gender and weight obfuscation}
    \label{fig:privacy_mobi_both}
\end{figure}

We first evaluate the obfuscation models on the MobiAct dataset when the user wishes to protect their gender. 
Figure~\ref{fig:privacy_mobi_gen} compares the intrusive inference accuracy obtained on the data obfuscated by each model. 
We consider five independent runs for each experiment, and plot the average and standard deviation (error bars) of accuracy over these runs.
Without data obfuscation, 
the adversary can easily predict the user's gender from the raw sensor data, attaining an average intrusive inference accuracy (F1 score) of $97.53\%$ ($97.50\%$).
The data obfuscated by ObscureNet yields an average intrusive inference accuracy (F1 score) of $52.04\%$ ($52.04\%$), 
which is close to random guessing, i.e. 50\%.
The Diffusion baseline shows the worst privacy loss with an average intrusive inference accuracy (F1 score) of $67.26\%$ ($66.33\%$). 
This suggests that without using the private attribute for negative conditioning, 
this baseline can provide only limited protection for the gender attribute due to the white-listing characteristic of diffusion models.
By introducing the negative conditioning, the Diffusion with Negation baseline drastically reduces the privacy loss and achieves an average intrusive inference accuracy (F1 score) of $49.72\%$ ($48.97\%$). 
Yet the improvement in privacy loss comes at the expense of reducing data utility (as discussed below), 
due to the entanglement of public and private attributes.
PrivDiffuser integrates negative conditioning along with mutual information-based regularization to achieve an average intrusive inference accuracy (F1 score) of $51.43\%$ ($50.49\%$), 
slightly underperforming Diffusion with Negation. 
But this accuracy is still close to random guessing and better than ObscureNet.
We attribute this to the different levels of
entanglement between each gender class and the public attribute, reducing the effect of negative conditioning on the private attribute class that has a stronger correlation with the public attribute.

Next, we study the impact of data obfuscation on data utility when the user wishes to protect their gender. 
The result is reported in Table~\ref{tab:utility_mobi_both}.
We find that all obfuscation models maintain relatively high data utility.
Notably, the Diffusion with Negation baseline results in the lowest utility despite achieving the lowest privacy loss, showing an average activity recognition accuracy (F1 score) of $97.07\%$ ($84.54\%$). 
We attribute this to using the negative condition to remove information about the private attribute without disentangling the public and private attributes.
ObscureNet yields an average HAR accuracy (F1 score) of $97.21\%$ ($84.99\%$), outperforming the Diffusion with Negation baseline.
The Diffusion baseline shows the highest HAR accuracy (F1 score) of $97.96\%$ ($87.44\%$), underscoring the outstanding capability of a conditional diffusion model to produce realistic data.
However, it fails to offer strong privacy protection as explained earlier.
PrivDiffuser yields the second-best data utility, with an average HAR accuracy (F1 score) of $97.40\%$ ($85.49\%$).
We conclude that PrivDiffuser yields the best privacy-utility trade-off in the gender obfuscation task, outperforming the state-of-the-art GAN-based obfuscation baseline.

\subsubsection{Protecting Weight}
\label{subsubsec:mobi_weight}
We now evaluate the obfuscation models when users want to protect their weight (a ternary attribute) on MobiAct. Figure~\ref{fig:privacy_mobi_weight} shows the result.
ObscureNet achieves a decent average intrusive inference accuracy (F1 score) of $38.88\%$ ($35.15\%$). 
Similar to the gender obfuscation case, the Diffusion baseline offers the worst privacy protection with an average weight inference accuracy (F1 score) of $55.86\%$ ($51.54\%$). 
However, this result is still encouraging as it reduces the intrusive inference accuracy by $35.88\%$ compared to raw sensor data,
without requiring knowledge of the private attribute.
Both the Diffusion with Negation baseline and PrivDiffuser outperform ObscureNet in terms of privacy loss, showing an intrusive inference accuracy (F1 score) of $36.61\%$ ($32.39\%$) and $35.78\%$ ($31.53\%$), respectively.
PrivDiffuser offers the best privacy protection since the weight inference accuracy is closest to the random guessing level, i.e. $33.33\%$.

Finally, we look at data utility in the weight obfuscation task.
As Table~\ref{tab:utility_mobi_both} shows, the Diffusion with Negation baseline achieves the lowest HAR accuracy ($94.93\%$) 
because of the application of negative conditioning without disentangling the attributes. 
ObscureNet and PrivDiffuser show nearly the same performance, achieving an average HAR accuracy (F1 score) of $97.02\%$ ($84.29\%$) and $97.03\%$ ($84.17\%$), respectively.
Specifically, the F1 score of ObscureNet is slightly higher, due to its higher accuracy in the walking upstairs activity. 
This is because ObscureNet trains a dedicated model for each activity, allowing each model to focus on a simpler task, which offers a slight advantage in maintaining data utility compared to PrivDiffuser.
Regardless, PrivDiffuser outperforms ObscureNet in terms of privacy loss by $3.10\%$ ($3.62\%$) w.r.t. the intrusive inference accuracy (F1 score). Overall, PrivDiffuser yields a better trade-off between privacy and utility.
In Appendix~\ref{appendix:B}, we further extend our evaluation to protecting multiple user attributes.

\begin{table}[t!]
    \centering
    \resizebox{\linewidth}{!}{
    \begin{tabular}{l c c c c c}
    \textbf{Model} & Downstairs & Upstairs & Walking & Jogging  & \textbf{Overall} (Accu./F1)\\
    \toprule
    Raw Data & 96.60\% & 94.40\% & 99.00\% & 97.40\% & 97.47\% / 96.59\% \\
    ObscureNet & 79.82\% & 95.28\% & 98.58\% & 97.22\% & 94.96\% / 93.05\% \\
    Diffusion & 96.10\% & 96.08\% & 95.68\% & 98.02\% & 96.28\% / 95.64\% \\
    Diffusion with Negation & 96.06\% & 96.34\% & 95.64\% & 97.94\% & 96.28\% / 95.64\% \\
    PrivDiffuser & 96.14\% & 96.44\% & 95.64\% & 98.00\% & 96.32\% / 95.69\% \\
    \bottomrule
   \end{tabular}}
    \caption{Average activity recognition accuracy on MotionSense  for gender obfuscation}
    \label{tab:utility_motion_gen}
\end{table}

\subsection{Case Study 2: HAR on MotionSense}
\begin{figure}[!t]
    \centering
    \begin{minipage}{.48\linewidth}
    \centering
    \includegraphics[width=\linewidth]{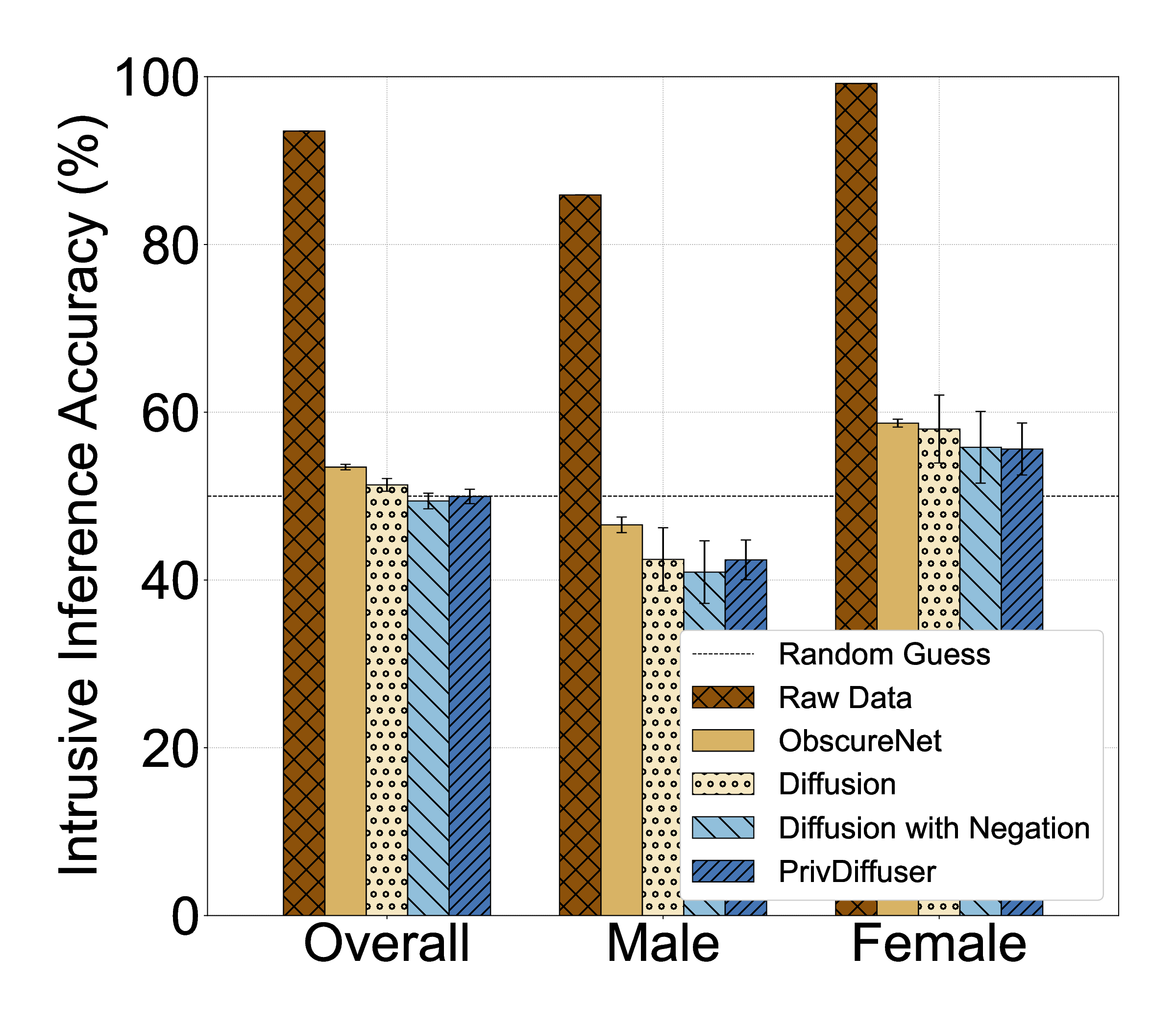}
    \caption{Intrusive inference accuracy on MotionSense for gender obfuscation}
    \label{fig:privacy_motion_gen}
    \end{minipage}%
    \hfill
    \hspace{2mm}
    \begin{minipage}{.48\linewidth}
        \centering
        \includegraphics[width=\linewidth]{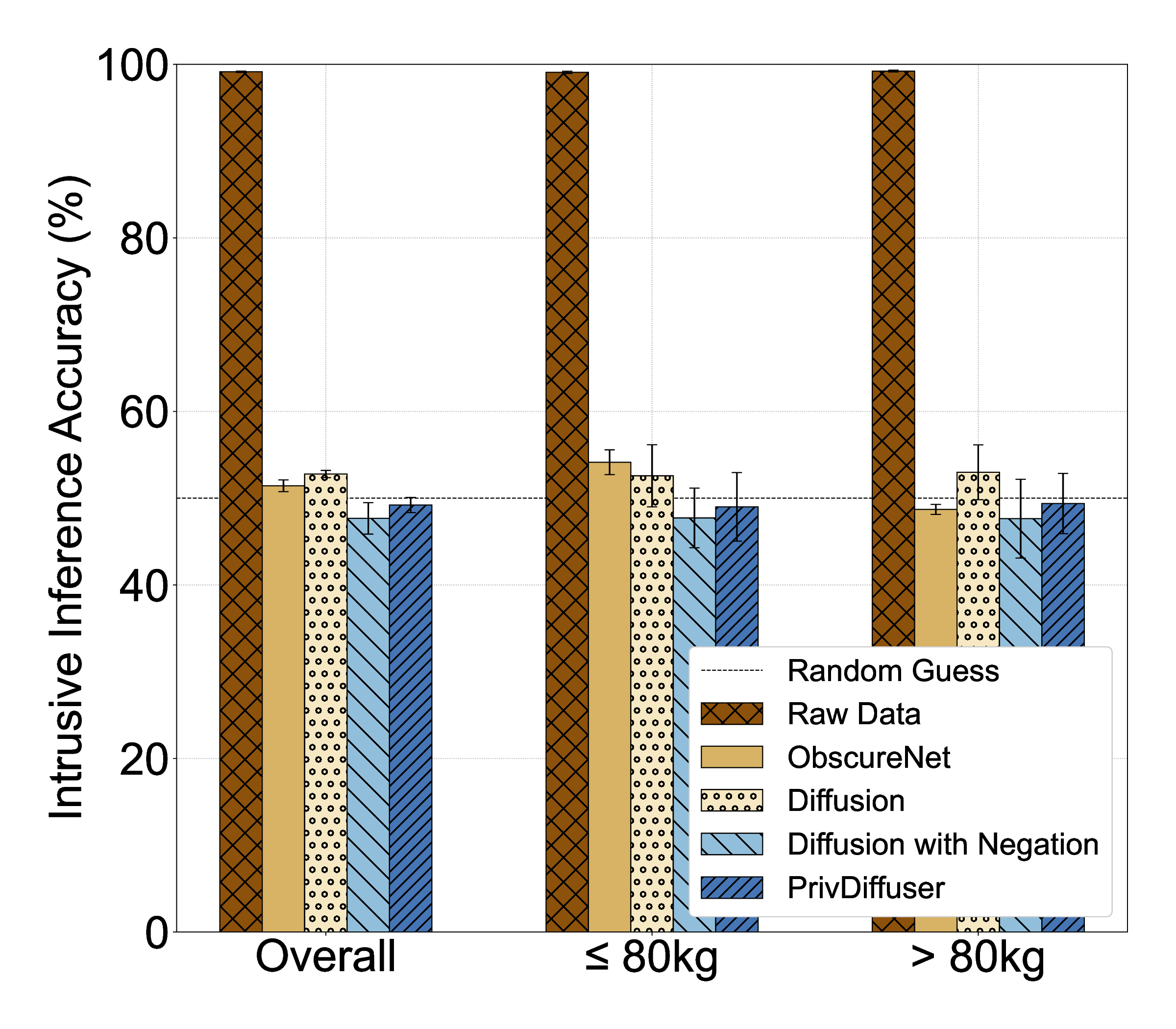}
        \caption{Intrusive inference accuracy on WiFi-HAR for weight obfuscation}
        \label{fig:privacy_wifi_weight}
    \end{minipage}
\end{figure}

We then evaluate the obfuscation models on the MotionSense dataset for gender obfuscation and report the intrusive inference accuracy in Figure~\ref{fig:privacy_motion_gen}.
Without applying data obfuscation, the intrusive inference model achieves a high average accuracy (F1 score) of $93.52\%$ ($93.26\%$) in predicting the gender attribute.
Data obfuscated by ObscureNet reduces the intrusive inference accuracy (F1 score) to $53.46\%$ ($52.61\%$). 
We find that the Diffusion baseline offers surprisingly good privacy-preserving performance and outperforms ObscureNet by only conditioning on the public attribute, showing an average intrusive inference accuracy (F1 score) of $51.34\%$ ($50.18\%$). 
Although the Diffusion baseline does not consistently outperform ObscureNet on all datasets, 
the white-listing characteristic could still offer promising privacy protection when the public and private attributes have relatively weak entanglement.
By incorporating the negative condition, the Diffusion with Negation baseline further reduces the intrusive inference accuracy (F1 score) to $49.42\%$ ($48.31\%$).
With mutual information-based regularization, PrivDiffuser 
yields an intrusive inference accuracy (F1 score) of $49.96\%$ ($49.00\%$).
For MotionSense, all diffusion-based %
models outperform the GAN-based baseline in terms of privacy loss.

We report the HAR accuracy when obfuscating gender on MotionSense in Table~\ref{tab:utility_motion_gen}.
In particular, ObscureNet yields good data utility, with an average HAR accuracy of $94.96\%$ ($93.05\%$), but underperforms all diffusion-based obfuscation models.
The Diffusion and Diffusion with Negation baselines achieve the HAR accuracy (F1 sore) of $96.28\%$ ($95.64\%$) on average.
PrivDiffuser achieves the highest average HAR accuracy (F1 score) of $96.32\%$ ($95.69\%$). 
In summary, for gender obfuscation in the MotionSense dataset, PrivDiffuser and the two diffusion-based baseline models outperform the GAN-based baseline in terms of privacy loss and data utility.

\subsection{Case Study 3: HAR on WiFi-HAR}
Lastly, we extend our evaluation study to obfuscate data collected using RF sensors on the WiFi-HAR dataset.
Figure~\ref{fig:privacy_wifi_weight} shows the intrusive inference accuracy on the WiFi-HAR dataset when obfuscating the binary weight attribute.
ObscureNet achieves an average weight inference accuracy (F1 score) of $51.43\%$ ($51.39\%$).
Although the Diffusion baseline achieves slightly higher (${\sim}1.35\%$) weight inference accuracy than ObscureNet, it still offers decent protection for the weight attribute, highlighting its white-listing characteristic.
The Diffusion with Negation baseline offers stronger privacy protection thanks to the negative conditioning, showing an average intrusive inference accuracy (F1 score) of $47.67\%$ ($47.62\%$).
PrivDiffuser yields an average intrusive inference accuracy (F1 score) of $49.21\%$ ($49.15\%$), which is only $0.79\%$ away from the ideal $50\%$ random guessing accuracy.

We evaluate the activity recognition accuracy on the obfuscated data and present the results in Table~\ref{tab:utility_wifi_weight}.
ObscureNet yields the lowest overall HAR accuracy (F1 score) of $86.37\%$ ($86.17\%$).
All diffusion-based obfuscation models yield nearly the same level of utility, outperforming ObscureNet.
Specifically, the Diffusion with Negation baseline achieves a lower HAR accuracy (F1 score) of $88.17\%$ ($88.11\%$).
PrivDiffuser outperforms ObscureNet w.r.t. the average HAR accuracy (F1 score) by $1.81\%$ ($1.95\%$). 

In summary, %
evaluation on %
three datasets indicates that PrivDiffuser consistently yields a superior privacy-utility trade-off, regardless of the sensing modality and %
user-specified private attribute.

\begin{table}[!t]
    \centering
    \resizebox{\linewidth}{!}{
    \begin{tabular}{l c c c c c}
    \textbf{Model} & Standing & Turning & Lying & Sitting  & \textbf{Overall} (Accu./F1)\\
    \toprule
    Raw Data & 95.77\% & 97.73\% & 99.40\% & 98.58\% & 97.88\% / 97.88\% \\
    ObscureNet & 86.82\% & 72.46\% & 96.68\% & 89.44\% & 86.37\% / 86.17\% \\
    Diffusion & 85.94\% & 96.64\% & 92.14\% & 78.38\% & 88.30\% / 88.24\% \\
    Diffusion with Negation & 85.76\% & 96.56\% & 91.78\% & 78.46\% & 88.17\% / 88.11\% \\
    PrivDiffuser & 85.76\% & 96.70\% & 91.56\% & 78.62\% & 88.18\% / 88.12\% \\
    \bottomrule
   \end{tabular}}
    \caption{Average activity recognition accuracy on WiFi-HAR dataset for weight obfuscation}
    \label{tab:utility_wifi_weight}
\end{table}

\subsection{Navigating Privacy-Utility Trade-offs}
\label{subsec:tuning_tradeoff}
We now turn our attention to how users with diverse privacy needs can navigate the privacy-utility trade-off.
We use the MotionSense dataset as an example and consider activity as the public attribute and gender as the private attribute.
To control the data utility and privacy, we adjust the hyperparameters $w_1$ and $w_2$  introduced in~(\ref{eq:negation_loss_ddim}), respectively.
Concretely, we assign one of the following four values to $w_2$: 0, 0.1, 0.2, and 0.3, 
where $w_2=0$ corresponds to the Diffusion baseline.
Then, for each $w_2$ value, 
we assign 9 different values from $\{1, {\cdots}, 9\}$ to $w_1$.
Note that a higher $w_1$ imposes stronger guidance on the positive condition (public attribute) 
which is expected to improve data utility.
Similarly, a higher $w_2$ imposes stronger guidance on the negative condition (private attribute), 
which is expected to improve privacy protection.
We study the effect of different values of $w_1$ and $w_2$ on the same pre-trained obfuscation model, 
meaning that these values are updated at the sampling stage 
to adjust the privacy-utility trade-off offered by PrivDiffuser. 
Thus, retraining the diffusion model, 
the surrogate utility model, and the auxiliary privacy model would not be necessary.
\textit{This is the main advantage of the proposed guidance approaches,
which sets PrivDiffuser apart from GAN-based obfuscation models.
}%

\begin{figure}[!t]
\centering
\includegraphics[width=\linewidth]{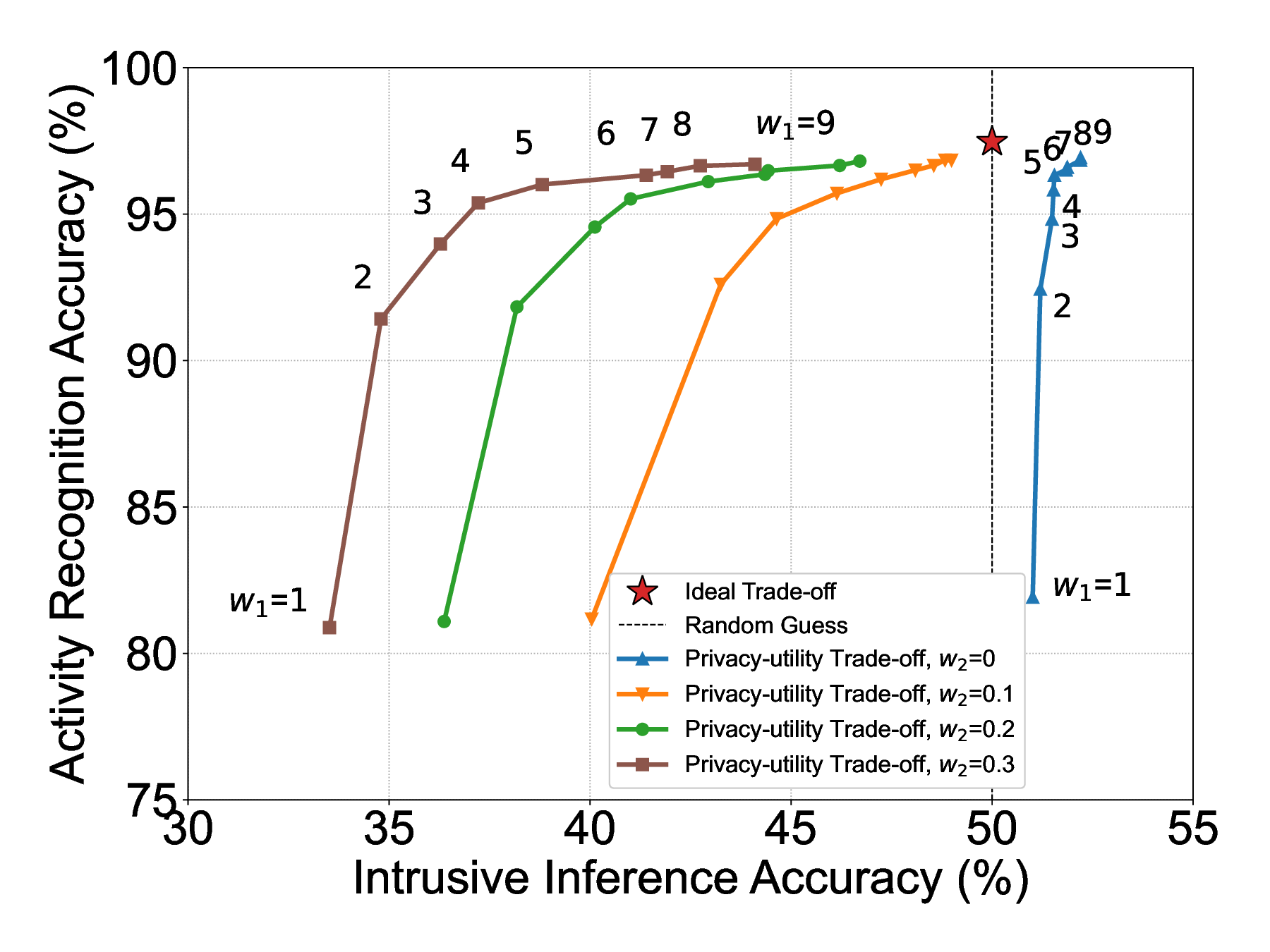}
\caption{Privacy-utility trade-off curves obtained by tuning the hyperparameters $w_1$ and $w_2$ of PrivDiffuser post training for gender obfuscation on MotionSense. Note that both axes are exaggerated. The activity recognition accuracy increases with $w_1$ in all curves, we annotate only two curves for brevity.}
\label{fig:trade_off}
\end{figure}

We illustrate the obtained trade-off curves in Figure~\ref{fig:trade_off}. 
The y-axis shows the activity recognition accuracy (a measure of utility), and the x-axis shows the gender inference accuracy (a measure of privacy loss). 
A red star is drawn to mark the ideal trade-off, which is defined as the HAR accuracy obtained on the raw sensor data ($97.47\%$) and the perfect random guessing accuracy of $50\%$ for the intrusive inference that involves the binary gender attribute.
For a fixed value of $w_2$, we find that increasing the value of $w_1$ substantially enhances the strength of the positive condition and improves data utility when $w_1$ is relatively small (${\leq}5$). 
However, as $w_1$ further increases, the control over the positive condition becomes insignificant and can counteract the effect of the negative condition, i.e., causing an increase in the intrusive inference accuracy when $w_2$ is fixed. 
We attribute this to the fact that the latent representation of the public attribute $z_p$ is extracted without applying information disentanglement techniques, 
hence a stronger guidance on the public attribute could add more sensitive information to $z_p$ due to the entanglement problem. 
On the other hand, for a fixed value of $w_1$,
increasing $w_2$ can effectively control the intrusive inference accuracy with very limited impact on data utility, thanks to the mutual information-based regularization that alleviates the entanglement of the public attribute with the latent representation of the private attribute $z_s$.
In practice, one can determine the values of $w_1$ and $w_2$ that yield a reasonable privacy-utility trade-off via sequential grid search, where $w_2$ can be first set to a default value (e.g., 0) to search for $w_1$ that achieves the best data utility, then $w_2$ can be tuned when $w_1$ is fixed at this value. 
The key takeaway is that we can effectively balance the effect of positive and negative conditions post-training by tuning the two knobs $w_1$ and $w_2$. This enables users to navigate the privacy-utility trade-off according to their privacy needs, which may even change over time.

\subsection{Protecting Multiple Private Attributes}

    We use the MobiAct dataset which contains more user attributes to evaluate the effectiveness of our approach for protecting multiple private attributes through negative conditioning. 
    We consider user activity as the public attribute, and 
    first evaluate PrivDiffuser when protecting two private attributes, namely gender and weight. We then extend our evaluation to protect three private attributes, with user ID being the third one.
    We re-use the pre-trained surrogate utility model, the diffusion model conditioned on the public attribute, and the auxiliary privacy model trained to predict gender as introduced in Section~\ref{subsec:mobiact}, and only incorporate a newly trained auxiliary privacy model that predicts the weight attribute for protecting two private attributes. We also train the third auxiliary privacy model that predicts the user ID of the 36 users included in the dataset.
    The hyperparameters are set empirically as follows: $w_1{=}2.2$, $w_2{=}0.3$, $w_3{=}0.6$ when protecting two private attributes, and $w_1{=}2.8$, $w_2{=}0.5$, $w_3{=}1.0$, $w_4{=}1.5$ when protecting three private attributes, with $w_2$ , $w_3$ and $w_4$ %
    controlling the importance of excluding information about gender, weight, and user ID in the obfuscated data, respectively.
    We follow the same experimental setup described in Section~\ref{sec:evaluation} and repeat the experiment 5 times to report the average HAR accuracy and intrusive inference accuracy.
    
    When protecting two private attributes, the data obfuscated by PrivDiffuser achieves average HAR accuracy (F1 score) of $95.64\%$ ($81.19\%$). 
    This shows that PrivDiffuser can protect multiple private attributes with little impact on utility.
    The average intrusive inference accuracy (F1 score) on the private attribute gender and weight group is respectively $49.67\%$ ($48.92\%$) and $35.30\%$ ($30.90\%$), deviating from the random guessing level by only $0.33\%$ and $1.97\%$, respectively. Compared to the privacy protection offered for the weight attribute due to the white-listing characteristic of diffusion models
    which is discussed in Appendix~\ref{subsec:whitelisting_characteristic}
    (i.e. when gender is used as the only private attribute), 
    protecting two attributes using negative conditioning further reduces the intrusive weight inference accuracy by $14.14\%$ 
    with very limited impact on utility.

    When using PrivDiffuser to protect three private attributes, the obfuscated data achieves average HAR accuracy (F1 score) of $93.77\%$ ($78.40\%$), a slight decrease of $1.87\%$ ($2.79\%$) compared to the case of protecting two private attributes.
    The average intrusive inference accuracy (F1 score) on gender, weight, and user ID is $50.35\%$ ($49.27\%$), $36.04\%$ ($31.05\%$), and $15.62\%$ ($15.57\%$), respectively.
    For the first two private attributes, the accuracy is close to the random guessing level ($50\%$ and $33.33\%$ respectively).
    For the user ID attribute, the ideal random guessing accuracy is $2.78\%$ given that there are 36 classes. For this attribute, although the intrusive inference accuracy on the obfuscated data is higher than random guessing, PrivDiffuser still reduces the intrusive inference accuracy by $79.50\%$ compared to raw data. More importantly, protecting the additional user ID attribute does not worsen the privacy loss for the other two private attributes.
    This result confirms that PrivDiffuser can be easily extended to protect multiple private attributes 
    without re-training the surrogate utility model or the diffusion model.

\subsection{Supporting Multiple Downstream Tasks by Conditioning on Multiple Public Attributes}
    
    We also use the MobiAct dataset to evaluate the effectiveness of conditioning up to three public attributes. 
    First, for the case of conditioning two public attributes, we consider user activity and gender as the public attributes, and weight group as the private attribute. For conditioning three public attributes, we use activity, gender, and weight as the public attributes, and user ID as the private attribute.
    The hyperparameters are set empirically as follows: $w_1{=}2.2$, $w_2{=}2.8$, $w_{p_1}{=}1.6$, $w_{p_2}{=}1.6$ 
    for two public attributes; 
    $w_1{=}5.8$, $w_2{=}0.5$, $w_{p_1}{=}0.4$, $w_{p_2}{=}0.4$, $w_{p_3}{=}0.4$ for three public attributes. 
    The remaining experimental setup is identical to the setup described in Section~\ref{sec:evaluation}. 
    When considering gender and weight as public attributes, we reuse the intrusive inference models trained for evaluation in Section~\ref{subsec:eval_metrics}
    as the desired inference models.
    We report the desired inference and intrusive inference accuracy averaged over 5 runs.

    We first evaluate conditioning two public attributes.
    Considering utility, we find that data obfuscated by PrivDiffuser achieves $90.90\%$ ($72.36\%$) average accuracy (F1 score) for activity recognition and $91.36\%$ ($91.30\%$) average accuracy (F1 score) for gender recognition. 
    Compared to performing inferences directly on raw sensor data, the HAR accuracy and gender recognition accuracy decreased by $7.90\%$ and $6.17\%$, respectively, which is the cost of protecting the private attribute.
    Next, we examine the privacy-preserving performance of PrivDiffuser for weight obfuscation when conditioned on two public attributes. The obfuscated data yields an average intrusive inference accuracy of $46.87\%$ on the ternary weight attribute, reducing the intrusive inference accuracy by $44.87\%$ compared to when raw data is released. 
    Recall that in Section~\ref{subsec:mobiact}, when PrivDiffuser was conditioned on one public attribute (activity) and weight was the private attribute, PrivDiffuser achieved an average intrusive inference accuracy of $35.78\%$. 
    Thus, conditioning on a second public attribute causes an increase in intrusive inference accuracy due to the stronger entanglement between weight and gender.

    Finally, we consider three public attributes. As for utility, the obfuscated data yields average desired inference accuracy (F1 score) of $92.14\%$ ($67.56\%$), $92.74\%$ ($92.61\%$), and $84.26\%$ ($82.56\%$) for activity, gender, and weight group, respectively. Comparing to the previous case of conditioning on two public attributes, we find that the desired inference accuracy for activity and gender slightly improves when adding a third public attribute (weight). We believe that this is because the third public attribute (weight) is entangled with the other two public attributes, causing more information to be included about them in the obfuscated data when PrivDiffuser is conditioned on all three public attributes.
    As for privacy protection, the obfuscated data yields average intrusive inference accuracy (F1 score) of $12.39\%$ ($10.91\%$) for the user ID attribute, a $82.73\%$ ($84.21\%$) decrease compared to the raw data.
    This result confirms that PrivDiffuser can effectively protect the weight attribute, 
    while allowing downstream applications to infer multiple public attributes with relatively high accuracy.

\subsection{Quantifying Privacy and Utility via MI}
To further verify PrivDiffuser's effectiveness in reducing privacy loss without significantly impacting data utility, we use an information theoretic approach. 
Specifically, we use MINE to estimate the mutual information between raw data and corresponding public/private attribute label, and then between obfuscated data and corresponding public/private attribute label. 
We perform data obfuscation on the MobiAct dataset as an example, and consider activity as the public attribute and gender as the private attribute. 
We fix $w_1{=}5.8$ for positive conditioning, and estimate the MI between the obfuscated data and their public/private attribute labels for $w_2 {\in} \{0, 0.5, 1.0, 1.5, 2.0, 2.5\}$, using a separate MINE for each $w_2$.

The estimated MI between raw data and private attribute is 0.683.
We observe that when $w_2{=}0$, i.e. without negative conditioning, 
the MI between obfuscated data and private attribute decreases to 0.6428. This reduction in privacy loss is attributed to the white-listing characteristic of PrivDiffuser.
As $w_2$ increases, the estimated MI further decreases to 0.4512 and 0.4696 when $w_2{=}0.5$ and $1.0$, respectively. This confirms that PrivDiffuser is effectively reducing the correlation between the obfuscated data and the private attribute. 
As $w_2$ further increases to 1.5, 2.0, and 2.5, we find that the estimated MI between the obfuscated data and gender increases to 0.5977, 0.6107, and 0.6226, respectively (see Appendix~\ref{appendix:C}).
For these values, the negative conditioning becomes too strong, 
causing the obfuscated data to include information about a wrong private attribute class (e.g. an opposite gender) rather than just obscuring information about the actual private attribute class. This increases MI.
Turning to utility, the estimated MI between the raw sensor data and the public attribute is 0.8850.
For the obfuscated data, the MI with the public attribute is maintained at a relatively stable level for all $w_2$ values, with a slight downward trend from 0.8767 when $w_2{=}0$, to 0.8426 when $w_2{=}0.5$ and 0.8105 when $w_2{=}1.0$. The MI remains around 0.82 for $w_2{\geq}1.5$.
These findings further substantiate the effectiveness of PrivDiffuser in obfuscating sensor data.

\section{Conclusion}
PrivDiffuser is the first data obfuscation technique to leverage conditional diffusion models, offering several key advantages over prior GAN-based approaches. Unlike the image synthesis task, where only one guidance method is typically used for conditioning, we address unique challenges in data obfuscation
through a deliberate, domain-aware design that incorporates two guidance methods.
Specifically, we applied classifier-free guidance for conditioning on public attributes and classifier guidance for private attributes, thereby enabling efficient and flexible adjustment of the privacy-utility trade-off. Additionally, we jointly optimized MINE and the auxiliary privacy model loss to alleviate the entanglement of public and private attributes.
Through evaluation on 3 HAR datasets containing data collected by different types of sensors, 
we corroborated that PrivDiffuser achieves a better privacy-utility trade-off than the state-of-the-art GAN-based obfuscation, showing increased data utility by up to $1.81\%$ and reduced privacy loss by up to $3.42\%$.
We extended PrivDiffuser to protect multiple private attributes and to support desired inference tasks.
Compared to GAN-based obfuscation models, 
PrivDiffuser offers a distinct advantage: \emph{the possibility of customizing the privacy-utility trade-off for each user during the sampling stage without model retraining.}
Together with its better obfuscation performance, this makes PrivDiffuser a better fit for real-world applications where users have diverse privacy needs.

\emph{Limitations.}
Similar to the previous work on sensor data obfuscation~\cite{chen2024mass,hajihassnai2021obscurenet,li2021deepobfuscator,liu2019privacy,malekzadeh2019mobile,raval2019olympus}, we have shown the efficacy of PrivDiffuser for defending against AIAs based on empirical evaluation, without providing a formal privacy guarantee. This is due to the inherent challenges of obscuring sensitive attributes that are not explicitly disclosed but can be inferred from the shared data, especially in a continual setting.
Differential privacy is not suitable for mitigating AIAs in the continual setting and under practical adversarial assumptions, 
and designing an obfuscation technique with theoretical guarantees remains an open problem in this setting. 
Another limitation of PrivDiffuser lies in the limited adoption of GPU-equipped IoT devices, primarily due to their high cost and energy consumption. Additionally, mobile machine learning platforms currently offer only limited support for backpropagation, which is crucial for implementing classifier guidance. 
However, we anticipate that these challenges will diminish in the near future, as GPUs continue to become more energy-efficient and affordable, and as on-device training and fine-tuning gain broader support in mobile platforms.
Lastly, PrivDiffuser requires users to explicitly specify their private attribute(s) and that 
at least partial disentanglement of public and private attributes is possible.
If these assumptions do not hold, a reasonable privacy-utility trade-off may not be achievable.

\emph{Future work.} There are a few alternative approaches for incorporating multiple positive and negative conditions simultaneously.
For instance, it is possible to train a dedicated surrogate utility model for each public attribute 
and apply the compositional conditioning techniques proposed in~\cite{liu2022compositional,huang2023composer}.
We aim to explore these approaches and compare them to the multi-task learning approach
adopted in PrivDiffuser.
We plan to dynamically adjust the progressive diffusion process 
to adapt PrivDiffuser's computational overhead to the given time budget.
This is crucial to achieve a satisfactory privacy-utility trade-off on resource-constrained devices.

\bibliographystyle{IEEEtran}
\bibliography{ref}

\begin{thebibliography}{10}
\providecommand{\url}[1]{#1}
\csname url@samestyle\endcsname
\providecommand{\newblock}{\relax}
\providecommand{\bibinfo}[2]{#2}
\providecommand{\BIBentrySTDinterwordspacing}{\spaceskip=0pt\relax}
\providecommand{\BIBentryALTinterwordstretchfactor}{4}
\providecommand{\BIBentryALTinterwordspacing}{\spaceskip=\fontdimen2\font plus
\BIBentryALTinterwordstretchfactor\fontdimen3\font minus \fontdimen4\font\relax}
\providecommand{\BIBforeignlanguage}[2]{{%
\expandafter\ifx\csname l@#1\endcsname\relax
\typeout{** WARNING: IEEEtran.bst: No hyphenation pattern has been}%
\typeout{** loaded for the language `#1'. Using the pattern for}%
\typeout{** the default language instead.}%
\else
\language=\csname l@#1\endcsname
\fi
#2}}
\providecommand{\BIBdecl}{\relax}
\BIBdecl

\bibitem{hajihassnai2021obscurenet}
O.~Hajihassnai, O.~Ardakanian, and H.~Khazaei, ``{ObscureNet}: Learning attribute-invariant latent representation for anonymizing sensor data,'' in \emph{Proceedings of the International Conference on Internet-of-Things Design and Implementation}, 2021, pp. 40--52.

\bibitem{malekzadeh2019mobile}
M.~Malekzadeh \emph{et~al.}, ``Mobile sensor data anonymization,'' in \emph{Proceedings of the International Conference on Internet-of-Things Design and Implementation}, 2019, pp. 49--58.

\bibitem{Purushwalkam2021}
S.~Purushwalkam \emph{et~al.}, ``Audio-visual floorplan reconstruction,'' in \emph{IEEE/CVF International Conference on Computer Vision}, 2021, pp. 1163--1172.

\bibitem{abadi2016deep}
M.~Abadi, A.~Chu, I.~Goodfellow, H.~B. McMahan, I.~Mironov, K.~Talwar, and L.~Zhang, ``Deep learning with differential privacy,'' in \emph{Proceedings of the 2016 ACM SIGSAC conference on computer and communications security}, 2016, pp. 308--318.

\bibitem{erlingsson2014rappor}
{\'U}.~Erlingsson, V.~Pihur, and A.~Korolova, ``Rappor: Randomized aggregatable privacy-preserving ordinal response,'' in \emph{Proceedings of the 2014 ACM SIGSAC Conference on Computer and Communications Security}, 2014, pp. 1054--1067.

\bibitem{zheng2020privacy}
X.~Zheng and Z.~Cai, ``Privacy-preserved data sharing towards multiple parties in industrial {IoTs},'' \emph{IEEE journal on selected areas in communications}, vol.~38, no.~5, pp. 968--979, 2020.

\bibitem{weggenmann2022dp}
B.~Weggenmann, V.~Rublack, M.~Andrejczuk, J.~Mattern, and F.~Kerschbaum, ``Dp-vae: Human-readable text anonymization for online reviews with differentially private variational autoencoders,'' in \emph{Proceedings of the ACM Web Conference 2022}, 2022, pp. 721--731.

\bibitem{HCNN2021}
A.~Al~Badawi \emph{et~al.}, ``Towards the alexnet moment for homomorphic encryption: {HCNN}, the first homomorphic {CNN} on encrypted data with {GPUs},'' \emph{IEEE Transactions on Emerging Topics in Computing}, vol.~9, no.~3, pp. 1330--1343, 2021.

\bibitem{raval2019olympus}
N.~Raval \emph{et~al.}, ``Olympus: Sensor privacy through utility aware obfuscation.'' \emph{Proceedings on Privacy Enhancing Technologies}, vol. 2019, no.~1, pp. 5--25, 2019.

\bibitem{yang2023blinder}
X.~Yang and O.~Ardakanian, ``Blinder: End-to-end privacy protection in sensing systems via personalized federated learning,'' \emph{ACM Transactions on Sensor Networks}, vol.~20, no.~1, pp. 1--32, 2023.

\bibitem{ho2022classifier}
J.~Ho and T.~Salimans, ``Classifier-free diffusion guidance,'' \emph{arXiv preprint arXiv:2207.12598}, 2022.

\bibitem{dhariwal2021diffusion}
P.~Dhariwal and A.~Nichol, ``Diffusion models beat {GANs} on image synthesis,'' \emph{Advances in Neural Information Processing Systems}, vol.~34, pp. 8780--8794, 2021.

\bibitem{bansal2023universal}
A.~Bansal \emph{et~al.}, ``Universal guidance for diffusion models,'' in \emph{Proceedings of the IEEE/CVF Conference on Computer Vision and Pattern Recognition}, 2023, pp. 843--852.

\bibitem{liu2019privacy}
S.~Liu \emph{et~al.}, ``Privacy adversarial network: representation learning for mobile data privacy,'' \emph{Proceedings of the ACM on Interactive, Mobile, Wearable and Ubiquitous Technologies}, vol.~3, no.~4, pp. 1--18, 2019.

\bibitem{li2021deepobfuscator}
A.~Li \emph{et~al.}, ``{DeepObfuscator}: Obfuscating intermediate representations with privacy-preserving adversarial learning on smartphones,'' in \emph{Proceedings of the International Conference on Internet-of-Things Design and Implementation}, 2021, pp. 28--39.

\bibitem{li2020tiprdc}
A.~Li \emph{et~al.}, ``{TIPRDC}: Task-independent privacy-respecting data crowdsourcing framework for deep learning with anonymized intermediate representations,'' in \emph{Proceedings of the 26th ACM SIGKDD International Conference on Knowledge Discovery \& Data Mining}, 2020, pp. 824--832.

\bibitem{lin2020using}
Z.~Lin \emph{et~al.}, ``Using {GANs} for sharing networked time series data: Challenges, initial promise, and open questions,'' in \emph{Proceedings of the ACM Internet Measurement Conference}, 2020, pp. 464--483.

\bibitem{chen2024mass}
Y.~Chen \emph{et~al.}, ``Mass: Multi-attribute selective suppression for utility-preserving data transformation from an information-theoretic perspective,'' in \emph{Proceedings of the 41st International Conference on Machine Learning}, 2024, pp. 6519--6538.

\bibitem{sohl2015deep}
J.~Sohl-Dickstein \emph{et~al.}, ``Deep unsupervised learning using nonequilibrium thermodynamics,'' in \emph{International Conference on Machine Learning}.\hskip 1em plus 0.5em minus 0.4em\relax PMLR, 2015, pp. 2256--2265.

\bibitem{ho2020denoising}
J.~Ho, A.~Jain, and P.~Abbeel, ``Denoising diffusion probabilistic models,'' \emph{Advances in Neural Information Processing Systems}, vol.~33, pp. 6840--6851, 2020.

\bibitem{song2020denoising}
J.~Song, C.~Meng, and S.~Ermon, ``Denoising diffusion implicit models,'' \emph{arXiv preprint arXiv:2010.02502}, 2020.

\bibitem{yang2023privacy}
X.~Yang and O.~Ardakanian, ``Privacy through diffusion: A white-listing approach to sensor data anonymization,'' in \emph{Proceedings of the 5th Workshop on CPS\&IoT Security and Privacy}, 2023, pp. 101--107.

\bibitem{nichol2021glide}
A.~Q. Nichol \emph{et~al.}, ``{GLIDE}: Towards photorealistic image generation and editing with text-guided diffusion models,'' in \emph{Proceedings of the 39th International Conference on Machine Learning}.\hskip 1em plus 0.5em minus 0.4em\relax PMLR, 2022, pp. 16\,784--16\,804.

\bibitem{preechakul2022diffusion}
K.~Preechakul \emph{et~al.}, ``Diffusion autoencoders: Toward a meaningful and decodable representation,'' in \emph{Proceedings of the IEEE/CVF Conference on Computer Vision and Pattern Recognition}, 2022, pp. 10\,619--10\,629.

\bibitem{radford2021learning}
A.~Radford \emph{et~al.}, ``Learning transferable visual models from natural language supervision,'' in \emph{International Conference on Machine Learning}.\hskip 1em plus 0.5em minus 0.4em\relax PMLR, 2021, pp. 8748--8763.

\bibitem{kim2022diffusionclip}
G.~Kim, T.~Kwon, and J.~C. Ye, ``Diffusionclip: Text-guided diffusion models for robust image manipulation,'' in \emph{Proceedings of the IEEE/CVF Conference on Computer Vision and Pattern Recognition}, 2022, pp. 2426--2435.

\bibitem{rombach2022high}
R.~Rombach \emph{et~al.}, ``High-resolution image synthesis with latent diffusion models,'' in \emph{Proceedings of the IEEE/CVF conference on computer vision and pattern recognition}, 2022, pp. 10\,684--10\,695.

\bibitem{suvorov2022resolution}
R.~Suvorov \emph{et~al.}, ``Resolution-robust large mask inpainting with fourier convolutions,'' in \emph{Proceedings of the IEEE/CVF winter conference on applications of computer vision}, 2022, pp. 2149--2159.

\bibitem{du2020compositional}
Y.~Du, S.~Li, and I.~Mordatch, ``Compositional visual generation with energy based models,'' \emph{Advances in Neural Information Processing Systems}, vol.~33, pp. 6637--6647, 2020.

\bibitem{liu2022compositional}
N.~Liu \emph{et~al.}, ``Compositional visual generation with composable diffusion models,'' in \emph{European Conference on Computer Vision}.\hskip 1em plus 0.5em minus 0.4em\relax Springer, 2022, pp. 423--439.

\bibitem{gandikota2023erasing}
R.~Gandikota \emph{et~al.}, ``Erasing concepts from diffusion models,'' in \emph{Proceedings of the IEEE/CVF International Conference on Computer Vision}, 2023, pp. 2426--2436.

\bibitem{dong2024towards}
P.~Dong \emph{et~al.}, ``Towards test-time refusals via concept negation,'' \emph{Advances in Neural Information Processing Systems}, vol.~36, 2024.

\bibitem{armandpour2023re}
M.~Armandpour \emph{et~al.}, ``Re-imagine the negative prompt algorithm: Transform {2D} diffusion into {3D}, alleviate {Janus} problem and beyond,'' \emph{arXiv preprint arXiv:2304.04968}, 2023.

\bibitem{deng2023fairness}
W.~Deng \emph{et~al.}, ``On fairness of medical image classification with multiple sensitive attributes via learning orthogonal representations,'' in \emph{International Conference on Information Processing in Medical Imaging}.\hskip 1em plus 0.5em minus 0.4em\relax Springer, 2023, pp. 158--169.

\bibitem{church2017word2vec}
K.~W. Church, ``Word2vec,'' \emph{Natural Language Engineering}, vol.~23, no.~1, pp. 155--162, 2017.

\bibitem{ronneberger2015u}
O.~Ronneberger, P.~Fischer, and T.~Brox, ``U-net: Convolutional networks for biomedical image segmentation,'' in \emph{Proceedings of the 18th International Conference on Medical Image Computing and Computer-Assisted Intervention}.\hskip 1em plus 0.5em minus 0.4em\relax Springer, 2015, pp. 234--241.

\bibitem{vaswani2017attention}
A.~Vaswani \emph{et~al.}, ``Attention is all you need,'' \emph{Advances in Neural Information Processing Systems}, vol.~30, 2017.

\bibitem{belghazi2018mutual}
M.~I. Belghazi \emph{et~al.}, ``Mutual information neural estimation,'' in \emph{International conference on machine learning}.\hskip 1em plus 0.5em minus 0.4em\relax PMLR, 2018, pp. 531--540.

\bibitem{chatzaki2016human}
C.~Chatzaki \emph{et~al.}, ``Human daily activity and fall recognition using a smartphone’s acceleration sensor,'' in \emph{International Conference on Information and Communication Technologies for Ageing Well and e-Health}.\hskip 1em plus 0.5em minus 0.4em\relax Springer, 2016, pp. 100--118.

\bibitem{baha2020dataset}
A.~Baha’A \emph{et~al.}, ``A dataset for wi-fi-based human activity recognition in line-of-sight and non-line-of-sight indoor environments,'' \emph{Data in Brief}, vol.~33, p. 106534, 2020.

\bibitem{unet-git}
``Unet implementation,'' \url{https://github.com/openai/guided-diffusion}, 2024 [Online], accessed in 2024.

\bibitem{huang2023composer}
L.~Huang \emph{et~al.}, ``Composer: Creative and controllable image synthesis with composable conditions,'' in \emph{Proceedings of the 40th International Conference on Machine Learning}.\hskip 1em plus 0.5em minus 0.4em\relax PMLR, Jul 2023, pp. 13\,753--13\,773.

\end{thebibliography}

\clearpage
\appendix
\section{White-listing Characteristic of PrivDiffuser}
\label{subsec:whitelisting_characteristic}
In this appendix, we revisit the white-listing characteristic of diffusion-based obfuscation models.
In diffusion-based obfuscation, information about the public attribute is 
preserved in the obfuscated data through positive conditioning, 
hence it is considered a white-listed attribute.
The white-listing characteristic of diffusion-based obfuscation means that a wide range of attributes, 
excluding the white-listed attribute(s), will be protected.
This stems from the nature of diffusion models that generate data from noise.
Without explicit guidance based on attributes that are not in the white list, 
the generated data would ideally contain information of arbitrary classes for those attributes, 
causing an adversary to obtain near-random guessing accuracy for the respective intrusive inferences.
However, the public attribute can be entangled with other attributes in practice, 
so applying the positive condition(s) could reveal information about other attributes and increase the privacy loss. %
In this case, users can specify the attribute they deem private and employ the proposed negative conditioning technique to provide strong privacy protection for that attribute.

Figure~\ref{fig:latent} sheds light on this white-listing characteristic of diffusion models, 
where not only the user-specified private attribute (area \textcircled{\raisebox{-0.9pt}{2}}+\textcircled{\raisebox{-0.9pt}{3}}) but also other attributes (area \textcircled{\raisebox{-0.9pt}{4}}) are protected, 
assuming their entanglement with the public attribute is not strong. 
We use the MobiAct dataset as an example to further investigate this characteristic in two cases:
\begin{enumerate}[nosep]
    \item No private attribute is specified by the user. Thus, all attributes apart from the public attribute, including gender and weight, might be protected due to the white-listing characteristic of diffusion models. We study this case using the Diffusion baseline.
    \item Gender is the only private attribute specified by the user. In this case, the weight attribute might also be protected due to the white-listing characteristic of diffusion models. We study this case on all obfuscation models.
\end{enumerate}

In the first case, the white-listing characteristic of the Diffusion baseline 
can offer moderate privacy protection while maintaining high data utility as discussed in the case studies.
In the MobiAct dataset, the Diffusion baseline yields the highest utility, achieving $97.96\%$ HAR accuracy.
Meanwhile, the white-listing characteristic reduces the intrusive inference accuracy on the binary gender and ternary weight attribute from $97.53\%$ to $67.26\%$ and from $91.74\%$ to $55.86\%$, respectively. 
Similar observations can be made for MotionSense and WiFi-HAR datasets, where the Diffusion baseline yields high utility. 
The overall intrusive inference accuracy on the binary gender attribute in MotionSense is reduced to $51.34\%$ (from $93.52\%$), and the intrusive inference accuracy on the binary weight attribute in WiFi-HAR is reduced to $52.78\%$ (from $99.14\%$).
We find that for both MotionSense and WiFi-HAR, 
the Diffusion baseline offers outstanding privacy protection, when compared to the random guessing level. 
We attribute this to the underlying data distribution in these two datasets, 
where the public and non-public attributes are relatively less entangled than MobiAct.
Overall, the white-listing characteristic of a basic diffusion-based obfuscation model provides moderate privacy protection for multiple non-public attributes.

\begin{table}[!t]
    \centering
    \resizebox{\linewidth}{!}{
    \begin{tabular}{l c c c c }
    \textbf{Model} & Group 0 & Group 1 & Group 2 & \textbf{Overall}\\
    \toprule
    Raw Data & 91.96\% & 89.80\% & 97.10\% & 91.74\%  \\
    ObscureNet & 100.00\% & 0.00\% & 0.00\% & 61.41\% \\
    Diffusion & 58.12\% & 56.44\% & 43.84\% & 55.86\% \\
    Diffusion with Negation & 54.28\% & 45.44\% & 37.48\% & 48.98\%  \\
    PrivDiffuser & 54.82\% & 46.00\% & 37.40\% & 49.44\%  \\
    \bottomrule
   \end{tabular}}
    \caption{Intrusive inference accuracy on the non-private weight attribute for MobiAct dataset. Obfuscation models are trained to protect the gender attribute.}
    \label{tab:whitelist_mobi}
\end{table}

Next, we study the second case. In Table~\ref{tab:whitelist_mobi}, we report the intrusive inference accuracy on the weight attribute when the obfuscation models are trained to protect gender on MobiAct.
We find that when the GAN-based ObscureNet baseline is trained to protect gender, 
the obfuscated data contains substantial information about the weight attribute. 
Specifically, ObscureNet achieves an average weight inference accuracy of $61.41\%$, 
which is significantly lower than the accuracy obtained on the raw data 
but noticeably higher than other diffusion-based obfuscation models.
Examining the result for each weight group, 
it becomes evident that all data segments obfuscated by ObscureNet are classified as weight group 0, 
the largest weight group representing $50\%$ of total users. 
This shows that the GAN-based obfuscation model fails to provide meaningful protection for attributes that were not explicitly black-listed by the user.
The Diffusion baseline, however, provides moderate privacy protection for the weight attribute, demonstrating better privacy protection performance than ObscureNet by not only reducing the overall intrusive inference accuracy but also lowering the inference accuracy for all weight groups in a relatively balanced manner.
By introducing the negative condition, the Diffusion with Negation baseline and PrivDiffuser further reduce the intrusive inference accuracy to around $49\%$. 
This implies that applying the negative condition of the gender attribute also contributes to the white-listing characteristic and improves the privacy loss on the weight attribute. 
We believe this is due to the entanglement between the weight and gender attribute in the latent space, 
as removing information about gender also helps with protecting the weight attribute.

In summary, we have shown that the white-listing characteristic of diffusion-based obfuscation models 
could provide privacy protection for a broad range of attributes other than the user-specified public and private attributes. 
Furthermore, the adoption of negative conditions could enhance this effect 
when the private attribute is entangled with unspecified attributes. 

\section{Deployment on IoT Devices: Time and Space Complexity}
\label{appendix:B}
In Section~\ref{sec:impl}, the MobiAct dataset was used in a case study to measure the inference time of PrivDiffuser on an NVIDIA RTX 2080 Ti GPU. We found that obfuscating a batch of 128 data segments takes about 11 seconds, including the inference time of the auxiliary models, i.e. approximately 88 ms for obfuscating one data segment. This is less than the interval between two consecutive data segments, which is 200 ms assuming a sampling rate of 50Hz and stride length of 10 samples. Thus, data obfuscation can be performed in real-time on this GPU.

\begin{table}[t!]
\centering
\resizebox{\linewidth}{!}{
\begin{tabular}{lccccc}
\textbf{\# Public-Private Attribute(s)} & 1-1 & 1-2 & 1-3 & 2-1 & 3-1 \\
\midrule
\textbf{Sampling Time/Batch (ms)} & 11309.17 & 11928.23 & 12506.83 & 11191.35 & 11333.39 \\
\textbf{Sampling Time/Segment (ms)} & 88.35 & 93.19 & 97.71 & 87.43 & 88.54 \\
\bottomrule
\end{tabular}}
\caption{Average sampling time for conditioning different combinations of public and private attributes using a single RTX 2080 Ti GPU, when performing gender obfuscation on MobiAct dataset. Each batch contains 128 data segments.}
\label{tab:latency}
\end{table}

\begin{table}[!t]
\centering
\resizebox{\linewidth}{!}{
\begin{tabular}{lcc}
\textbf{Specification} & \textbf{NVIDIA RTX 2080 Ti} & \textbf{NVIDIA Jetson Orin NX (16GB)} \\
\toprule
Architecture & Turing (12 nm) & Ampere (8 nm) \\
CUDA Cores & 4,352 & 1,024 \\
Tensor Cores & 544 & 32 \\
Memory & 11 GB GDDR6 (352‑bit) & 16 GB LPDDR5 (128‑bit) \\
AI Performance & 228 TOPS (INT8) & 157 TOPS (INT8) \\
Power (TDP) & 250 W & 10–25 W \\
\bottomrule
\end{tabular}}
\caption{Specification comparison of the NVIDIA RTX 2080 Ti vs. NVIDIA Jetson Orin NX.}
\label{tab:jetson_compare}
\end{table}

We now investigate how the computational overhead could scale with more public or private attributes.
The time and space complexity of training PrivDiffuser, which needs to happen once, is expected to grow linearly with the number of private attributes because a dedicated auxiliary privacy model with a MINE network must be trained per private attribute. This is assuming that the overhead of training each auxiliary privacy model (and each MINE) is roughly the same.
At the sampling stage, the trained networks must be used for guiding the diffusion model via classifier guidance. Hence, the inference cost is also expected to increase linearly with the number of private attributes. 
When it comes to supporting multiple public attributes, we have used  multi-task learning to train a single surrogate utility model with multiple classification heads to extract latent representations, 
so the dimension of the latent representation will not scale with the number of public attributes. 
As a result, the increase in time and space complexity is expected to be rather small with more public attributes, for both training and sampling. 
In Table~\ref{tab:latency}, we show the sampling time of PrivDiffuser when performing data obfuscation on the MobiAct dataset with up to three public or private attributes, using a single RTX 2080 Ti GPU. 
We measure the sampling time per batch for at least 50 batches and report the average sampling time per batch and per data segment. 
The result confirms the above analysis.
Specifically, conditioning on two private attributes increases the sampling time per data segment by an average of 4.84 ms compared to conditioning on one private attribute, and adding a third private attribute further increases the sampling by an average of 4.52 ms; this is a near-linear increase in the sampling cost.
However, when increasing the number of conditioned public attributes, the average sampling time per data segment remains at around 88 ms. This shows that increasing the number of public attributes will have almost no impact on the sampling cost.

As a point of comparison, ObscureNet, our GAN-based baseline, 
supports protecting multiple private attributes by training multiple discriminator networks, but cannot incorporate multiple public attributes.
The obfuscation process in ObscureNet involves running each discriminator network. 
Therefore, the training and sampling overhead of ObscureNet is expected to grow almost linearly with the number of conditioned private attributes, assuming that the discriminators have a similar architecture. 
Since ObscureNet requires training a dedicated CVAE for each class of the public attribute, extending it to protect multiple private attributes would result in substantially higher training costs compared to PrivDiffuser. Additionally, in ObscureNet and other GAN-based approaches, changing the definition of the private attribute(s) requires re-training the entire obfuscation model (i.e. CVAE with discriminators).

Given that the classifier guidance technique requires computing gradients during sampling and existing mobile machine learning frameworks (e.g., PyTorch Mobile, ExecuTorch) have very limited support for back-propagation, it is challenging to evaluate the real-time latency of PrivDiffuser on mobile devices.
To get a sense of the overhead of running PrivDiffuser on GPU-equipped IoT/edge devices, in Table~\ref{tab:jetson_compare}, we compare our GPU (RTX 2080 Ti) with Jetson Orin NX that is designed for IoT/edge applications and can provide up to 157 sparse INT8 Tera Operations per Second (TOPS) with 16GB of memory. 
It can be seen that Jetson has more memory but handles $30\%$ less operations per second, suggesting that the obfuscation latency on GPU-equipped IoT devices will be slightly higher but in the same ballpark as our GPU.
Thus, we posit that running PrivDiffuser on IoT devices will become feasible as such devices are increasingly equipped with powerful GPUs.
It is worth mentioning that the computational overhead of PrivDiffuser mainly stems from the multiple steps of sampling and the relatively large dimension of sensor data segments. Hence the deployment on mobile/edge devices can benefit from further optimization, such as reducing the number of sampling steps or applying few-shot sampling techniques for diffusion models. Furthermore, applying latent diffusion models can drastically reduce the size of the sampled data. Data quantization and model pruning are expected to further improve the running time on resource-constrained devices.

In terms of the model size, among the three datasets, the largest diffusion model is the one trained for the WiFi-HAR dataset with a size of 282 MB, and each auxiliary classifier is around 495 MB. 
The MobiAct dataset has the smallest diffusion model with a size of 18 MB, with each classifier having a size of around 35 MB. 
Nevertheless, all three versions of PrivDiffuser would fit in the memory of GPU-equipped IoT/edge devices such as Jetson Orin NX.

\section{Estimating Mutual Information Between Obfuscated Data and Attributes}
\label{appendix:C}
The figure below compares the mutual information between obfuscated data and public/private attributes with the mutual information between raw sensor data and these attributes. In each case, the mutual information is estimated using a MINE network.

\begin{figure}[!t]
\centering
\includegraphics[width=\linewidth]{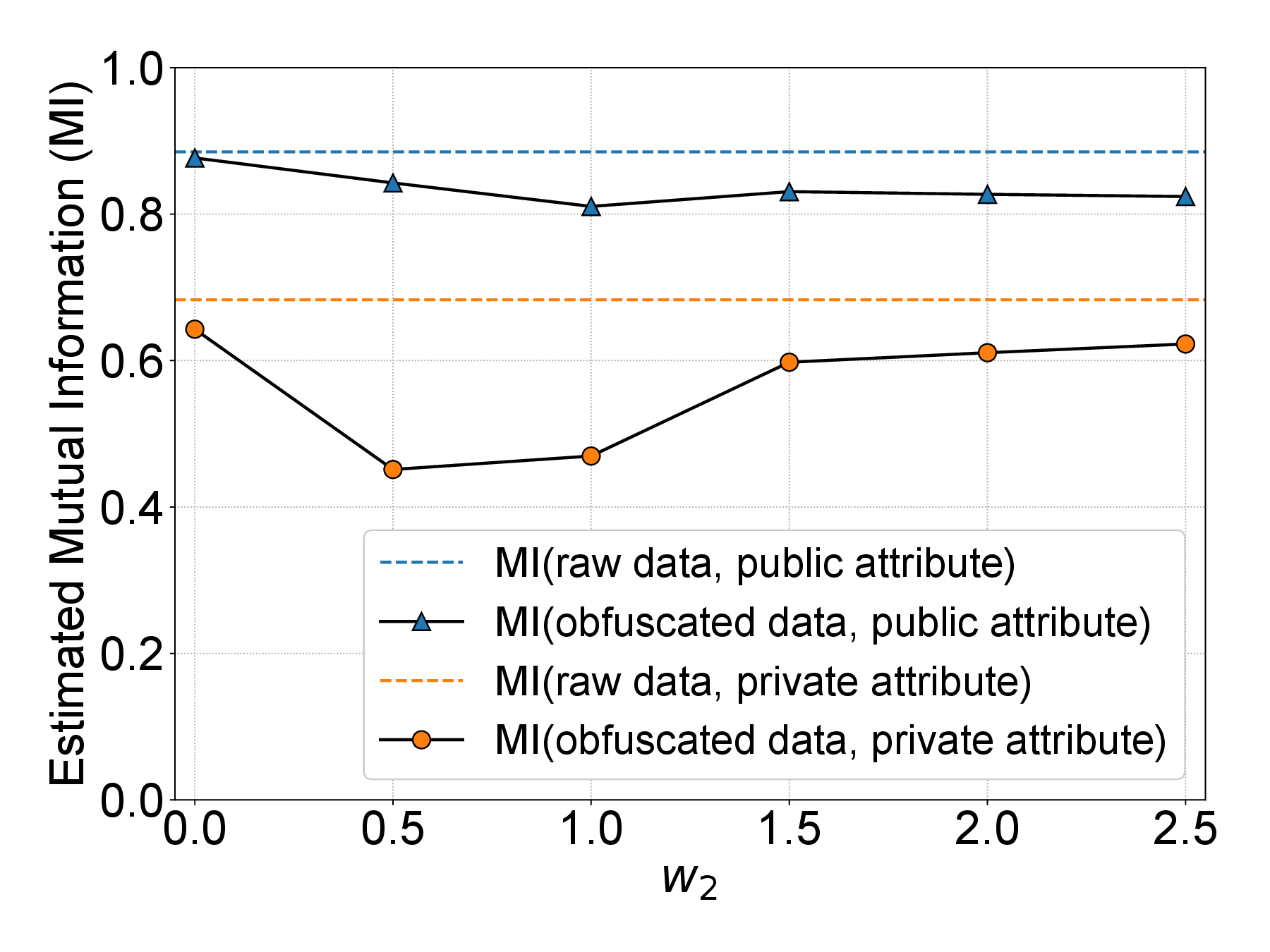}
\caption{Estimated MI between obfuscated data and public/private attribute when performing gender obfuscation on MobiAct dataset with various $w_2$ values and a fixed $w_1{=}5.8$.}
\label{fig:mutual_info}
\end{figure}

\end{document}